\documentclass[aps,prb,superscriptaddress,twocolumn]{revtex4-2}
\usepackage{amsmath, amssymb}
\usepackage{braket}
\usepackage{color}
\usepackage{graphicx}
\usepackage{ulem}
\usepackage{siunitx}

\begin{document}
\date{\today}
%%%%%%%%%%%%%%%%%%%%%%%%%%%%%%%%%%%%%%%%%%%%%%%%%%%%
\title{Local density of states as a probe for tunneling magnetoresistance effect: application to ferrimagnetic tunnel junctions}
\author{Katsuhiro Tanaka}
\affiliation{Research Center for Advanced Science and Technology, University of Tokyo, Komaba, Meguro-ku, Tokyo 153-8904, Japan}
\author{Takuya Nomoto}
\affiliation{Research Center for Advanced Science and Technology, University of Tokyo, Komaba, Meguro-ku, Tokyo 153-8904, Japan}
\author{Ryotaro Arita}
\affiliation{Research Center for Advanced Science and Technology, University of Tokyo, Komaba, Meguro-ku, Tokyo 153-8904, Japan}
\affiliation{Center for Emergent Matter Science, RIKEN, Wako, Saitama 351-0198, Japan}
%%%%%%%%%%%%%%%%%%%%%%%%%%%%%%%%%%%%%%%%%%%%%%%%%%%%
\begin{abstract}
We investigate the tunneling magnetoresistance (TMR) effect using the lattice models which describe the magnetic tunnel junctions (MTJ).
First, taking a conventional ferromagnetic MTJ as an example, 
we show that the product of the local density of states (LDOS) at the center of the barrier traces the TMR effect qualitatively.
The LDOS inside the barrier has the information on the electrodes and the electron tunneling through the barrier,
which enables us to easily evaluate the tunneling conductance more precisely than the conventional Julliere's picture.
We then apply this method to the MTJs with collinear ferrimagnets and antiferromagnets.
We find that the TMR effect in the ferrimagnetic and antiferromagnetic MTJs changes depending on the interfacial magnetic structures originating from the sublattice structure, 
which can also be captured by the LDOS.
Our findings will reduce the computational cost for the qualitative evaluation of the TMR effect,
and be useful for a broader search for the materials which work as the TMR devices showing high performance.
\end{abstract}
%%%%%%%%%%%%%%%%%%%%%%%%%%%%%%%%%%%%%%%%%%%%%%%%%%%%
\maketitle
%%%%%%%%%%%%%%%%%%%%%%%%%%%%%%%%%%%%%%%%%%%%%%%%%%%%
\section{Introduction}
\label{sec:introduction}
Utilizing the close connection between the spin and charge degrees of freedom of electrons in solids, 
spintronics has developed various phenomena that are novel from the viewpoint of fundamental physics and promising for industrial use ~\cite{Zutic2004_RevModPhys_76_323,Felser2007_AngewChemIntEd_46_668,Chappert2007_NatMater_6_813,Bader2010_AnnuRevCondensMatterPhys_1_71,Hirohata2020_JMagnMagnMater_509_166711}.
Among those, the tunneling magnetoresistance (TMR) effect~\cite{Julliere1975_PhysLettA_54A_225,Stearns1977_JMagnMagnMater_5_167} is one of the representative phenomena in its wide application ~\cite{Tsymbal2003_JPhysCondensMatter_15_R109,Zhang2003_JPhysCondensMatter_15_R1603,Ito2006_JMagnSocJpn_30_1,Yuasa2007_JPhysD_40_R337,Butler2008_SciTechnolAdvMater_9_014106}.
The TMR effect is observed in the magnetic tunnel junction (MTJ),
which consists of two magnetic electrodes and the insulating barrier in between.
The electrons can tunnel through the MTJ as a quantum mechanical current, 
and the tunneling resistances become different when the magnetic moments of the two electrodes align parallelly or antiparallelly.
The set of these two alignments with different tunneling resistances corresponds to a bit taking a binary 0 or 1,
which has been utilized to the magnetic head and the magnetic random access memory devices for the storages and the readout.
As well as the theoretical approaches~\cite{Slonczewski1989_PhysRevB_39_6995,Mathon1997_PhysRevB_56_11810,Butler2001_PhysRevB_63_054416,Mathon2001_PhysRevB_63_220403}, 
large TMR ratios have been experimentally observed in the MTJs such as the Fe(Co)/$\mathrm{Al_{2}O_{3}}$/Fe(Co)~\cite{Miyazaki1995_JMagnMagnMater_139_L231,Moodera1995_PhysRevLett_74_3273}, Fe(Co)(001)/MgO(001)/Fe(Co)~\cite{Yuasa2004_NatMater_3_868,Parkin2004_NatMater_3_862} and CoFeB/MgO/CoFeB systems~\cite{Djayaprawira2005_ApplPhysLett_86_092502,Ikeda2008_ApplPhysLett_93_082508}.
Ferromagnetic Heusler compounds have also been utilized as the electrodes thanks to their half-metalicity~\cite{Tanaka1999_JApplPhys_86_6239,Inomata2003_JpnJApplPhys_42_L419,Kammerer2004_ApplPhysLett_85_79,Liu2012_ApplPhysLett_101_132418}.
\par
While the main target of the spintronics was ferromagnets, 
recent spintronics has been extended to antiferromagnets and ferrimagnets owing to their superiorities to ferromagnets; the smaller stray field and the faster spin dynamics~\cite{MacDonald2011_PhilTransRSocA_369_3098,Gomonay2014_LowTempPhys_40_17,Jungwirth2016_NatNanotechnol_11_231,Baltz2018_RevModPhys_90_015005,Zelezny2018_NatPhys_14_220,Siddiqui2020_JApplPhys_128_040904,Fukami2020_JApplPhys_128_070401,Barker2021_JPhysSocJpn_90_081001,Kim2022_NatMater_21_24}.
The antiferromagnetic version of the spintronic phenomena, e.g.,
the giant magnetoresistance effect~\cite{Nunez2006_PhysRevB_73_214426,Saidaoui2014_PhysRevB_89_174430,Ghosh2022_PhysRevLett_128_097702} and the anomalous Hall effect~\cite{Chen2014_PhysRevLett_112_017205,Nakatsuji2015_Nature_527_212,Kiyohara2016_PhysRevApplied_5_064009,Nayak2016_SciAdv_2_e1501870},
has been developed.
Along with these advances, the TMR effect using antiferromagnets has also been intensively investigated ~\cite{Merodio2014_ApplPhysLett_105_122403,Jeong2016_NatCommun_7_10276,Saidaoui2017_PhysRevB_95_134424,Shao2021_NatCommun_12_7061,Smejkal2022_PhysRevX_12_011028,Dong2022_PhysRevLett_128_197201}.
While most of the studies have been theoretical attempts, 
experiments have also been developed;
the TMR effect is observed in the MTJ whose two electrodes are the ferromagnet and the ferrimagnetic Heusler compound~\cite{Jeong2016_NatCommun_7_10276}.
However, for more practical application of the MTJs with antiferromagnets and ferrimagnets to the devices, 
we should search for materials constructing the MTJs which show a large TMR ratio,
and handy methods for the search are required.
\par
In this paper, we examine the TMR effect using the lattice models mimicking the MTJs whose electrodes are made of collinear ferrimagnets, including the antiferromagnets.
Motivated by the studies indicating that the interfacial electronic structures affect the TMR effect and they can be probed by the local density of states (LDOS)~\cite{Tsymbal2005_JApplPhys_97_10C910,Tsymbal2007_ProgMaterSci_52_401,Masuda2020_PhysRevB_101_144404,Masuda2021_PhysRevB_103_064427,Masuda2021_PhysRevB_104_L180403},
we particularly focus on the LDOS to analyze the TMR effect.
We find that the product of the LDOS at the center of the barrier usually reproduces the transmission properties qualitatively in the ferrimagnetic MTJs as well as the ferromagnetic ones.
The LDOS has the information both on the magnetic properties of electrodes and on the tunneling electrons.
Besides, from the physics point of view, 
we show that multiple configurations can be realized in the ferrimagnetic MTJs due to the sublattice structure for each of the parallel and antiparallel magnetic configurations.
The resultant TMR effect changes depending on the configurations,
which suggests that the magnetic configurations should be carefully examined when we deal with the ferrimagnetic MTJs.
\par
\begin{figure}[t]
	\centering
	\includegraphics[width=86mm]{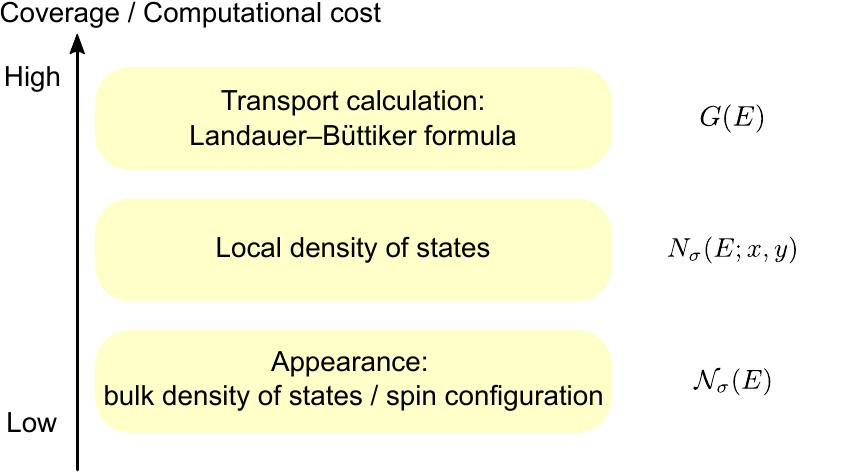}
	\caption{%
		Hierarchy for the evaluation of the tunneling magnetoresistance effect.
		The upper has a broader coverage and gives more quantitative results,
		with the higher calculation cost.
		Symbols in the right side, $G(E)$, $N_{\sigma}(E; x, y)$, and $\mathcal{N}_{\sigma}(E)$, 
		denote the conductance, the local density of states, and the density of states of the bulk, respectively,
		which are the quantities obtained in each calculation.
	}
	\label{fig:concept}
\end{figure}

Considering the above qualitative estimation in terms of the LDOS,
we present a hierarchy for evaluating the TMR effect in Fig.~\ref{fig:concept}.
To quantitatively estimate the TMR effect, 
we have to calculate the conductance itself through the Landauer--B\"{u}ttiker formula~\cite{Landauer1957_IBMJResDev_1_3,Landauer1970_PhilMag_21_863,Buttiker1986_PhysRevLett_57_1761,Buttiker1988_IBMJResDevelop_32_317}.
Technically, this method can be applied to any system and gives us highly accurate results, 
whereas its numerical cost is often expensive, particularly in calculating from first-principles. 
By contrast, the Julliere's picture, 
which claims that the density of states of the bulk electrodes determines the efficiency of the MTJ~\cite{Julliere1975_PhysLettA_54A_225,Maekawa1982_IEEETranMagn_18_707},
is a simple and convenient picture to predict the TMR effect.
However, the picture is valid only for limited cases, 
and currently it is found that the electronic states of the tunneling electrons are significant rather than those of the bulk electrodes.
Our results can be placed between these two methods.
Calculating the LDOS of the MTJs is much easier than the Landauer--B\"{u}ttiker calculation.
Additionally, the estimation from the LDOS can cover a broader range of the MTJs with higher reliability than the prediction from the electronic structures of the bulk electrodes.
\par
The remainder of this paper is organized as follows.
In Sec.~\ref{sec:method}, we introduce the model describing the MTJ.
We simulate the TMR effect using the ferromagnetic electrodes in Sec.~\ref{sec:ferromagnetic_TMR},
and see how the LDOS works for predicting the tunneling conductance.
In Sec.~\ref{sec:ferrimagnetic_TMR}, we calculate the TMR effect in the ferrimagnetic and antiferromagnetic MTJs applying the prediction from the LDOS.
We discuss in Sec.~\ref{sec:remarks} the hierarchy shown in Fig.~\ref{fig:concept} and the correspondence between our models and the MTJ with real materials.
Section~\ref{sec:summary} is devoted to the summary and perspective of this study.
%%%%%%%%%%%%%%%%%%%%%%%%%%%%%%%%%%%%%%%%%%%%%%%%%%%%
%%%%%%%%%%%%%%%%%%%%%%%%%%%%%%%%%%%%%%%%%%%%%%%%%%%%
\section{Model and Method}
\label{sec:method}
\begin{figure}[t]
	\centering
	\includegraphics[width=86mm]{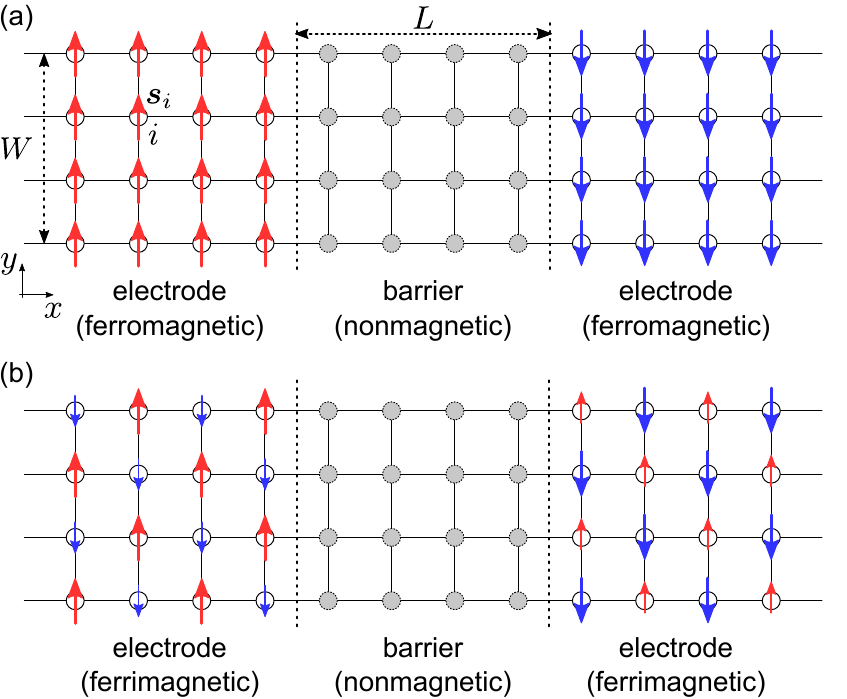}
	\caption{%
		Schematics of the two-dimensional magnetic tunnel junctions (MTJ) used in our calculations.
		(a) MTJ with the ferromagnetic electrodes.
		(b) MTJ with the ferrimagnetic electrodes.
		Arrows represent the localized spin moments on the electrodes.
	}
	\label{fig:MTJ_ferro_ferri}
\end{figure}

We construct the two-dimensional square lattice MTJ using two semi-infinite lattices which work as electrodes and the barrier in between,
which is schematically shown in Fig.~\ref{fig:MTJ_ferro_ferri}.
We treat the tight-binding Hamiltonian with the $s$--$d$ coupling on this system, 
which is given as
\begin{align}
	\mathcal{H}
& = \mathcal{H}_{0} + \mathcal{H}_{t} + \mathcal{H}_{s-d}, \label{eq:hamiltonian_all} \\
	\mathcal{H}_{0}
& = \sum_{i} \varepsilon_{i} n_{i}, \label{eq:hamiltonian_0} \\
	\mathcal{H}_{t}
& = - t \sum_{\braket{i, j}, \sigma} \left( c_{i, \sigma}^{\dag} c_{j, \sigma} + \text{h.c.} \right), \label{eq:hamiltonian_t} \\
	\mathcal{H}_{s\text{--}d}
& = - J \sum_{i \in \text{electrode}} \left( \boldsymbol{s}_{i} \cdot \boldsymbol{\sigma} \right)^{\alpha\beta} c_{i, \alpha}^{\dag} c_{i, \beta}. \label{eq:hamiltonian_sd}
\end{align}
Here, $c_{i, \sigma}^{\dag}$ ($c_{i, \sigma}$) is the creation (annihilation) operator of an electron with spin-$\sigma$ on the $i$-th site, 
and $n_{i} = \sum_{\sigma} c_{i, \sigma}^{\dag} c_{i, \sigma}$ is the number operator.
The on-site potential is denoted as $\varepsilon_{i}$,
and the electron hopping between two sites is written as $t$.
The summation in $\mathcal{H}_{t}$ is taken over the neighboring two sites,
which is expressed by $\braket{i, j}$.
The effect of the magnetism of the electrodes is introduced by the $s$--$d$ coupling, $\mathcal{H}_{s\text{--}d}$, 
where the localized spin moment, $\boldsymbol{s}_{i}$,
and the conducting electrons couple each other with the magnetic interaction constant, $J$.
The spin degrees of freedom of the conducting electrons are expressed by $\boldsymbol{\sigma}$, 
which is the vector representation of the $2\times 2$ Pauli matrices.
We take the two-dimensional cartesian coordinate, $(x, y)$, for the MTJ,
where the $x$-axis is parallel to the conducting path which infinitely extends, 
and the $y$-axis is perpendicular to the conducting path (see Fig.~\ref{fig:MTJ_ferro_ferri}(a)).
The width of the barrier in the $x$-direction is $L$,
and the width of the MTJ in the $y$-direction is $W$.
The lattice constant is taken to be unity.
We hereafter impose the open boundary condition on the $y$-direction,
while we confirmed that the periodic boundary condition does not change the overall results.
\par
The calculations of the transmissions are performed using the \textsc{kwant} package~\cite{Groth2014_NewJPhys_16_063065}, 
in which the quantum transport properties are computed based on the scattering theory and the Landauer--B\"{u}ttiker formula.
%%%%%%%%%%%%%%%%%%%%%%%%%%%%%%%%%%%%%%%%%%%%%%%%%%%%
%%%%%%%%%%%%%%%%%%%%%%%%%%%%%%%%%%%%%%%%%%%%%%%%%%%%
\section{Tunneling magnetoresistance effect with ferromagnetic electrodes}
\label{sec:ferromagnetic_TMR}
%%%%%%%%%%%%%%%%%%%%%%%%%%%%%%%%%%%%%%%%%%%%%%%%%%%%
\begin{figure*}[t]
	\centering
	\includegraphics[width=172mm]{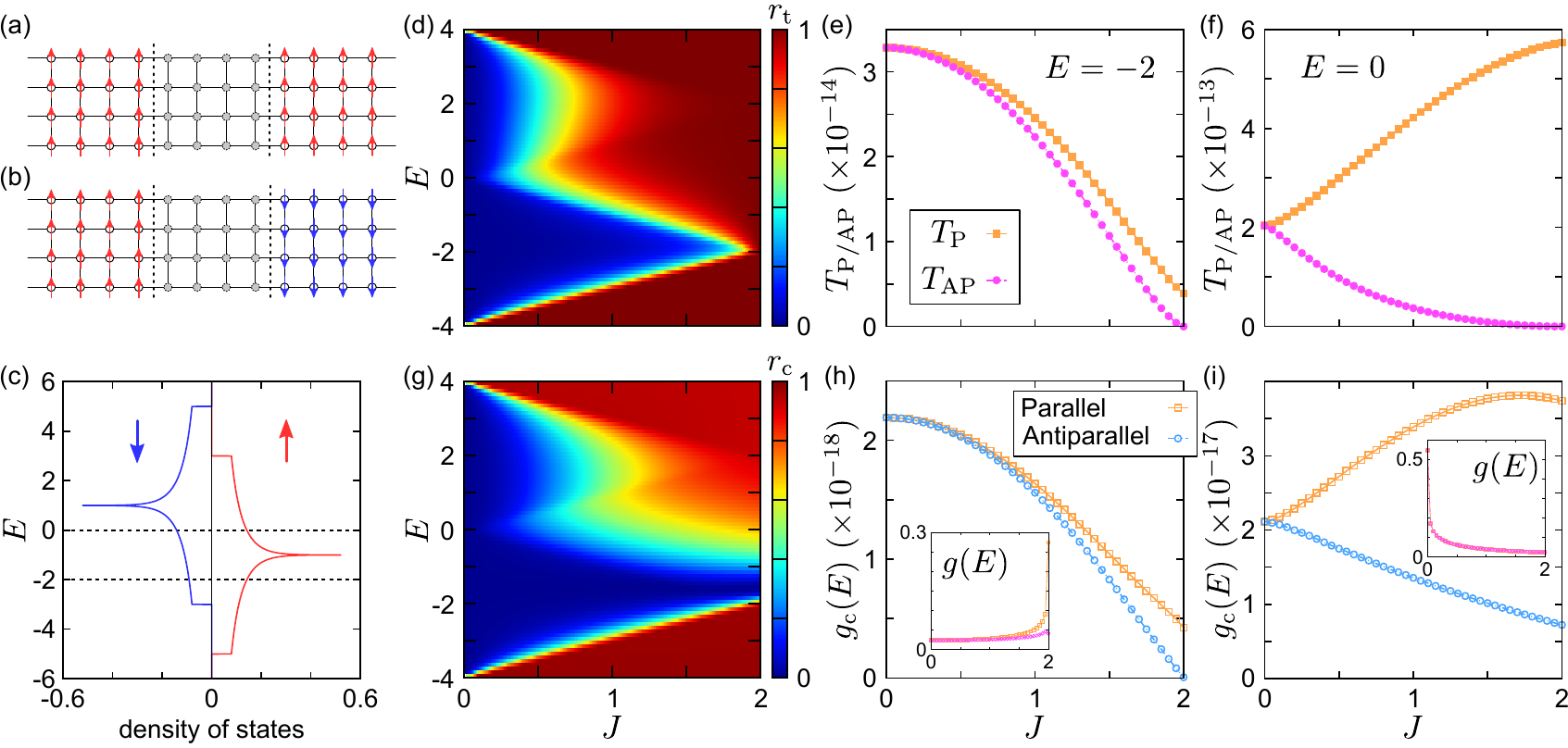}
	\caption{%
		(a), (b)
		Schematic pictures of (a) the parallel and (b) antiparallel configuration of the magnetic tunnel junction (MTJ) with ferromagnetic electrodes.
		(c) Spin-resolved density of states of the electrode with $J = 1$.
		Broken lines are the energies where we show the results in (e), (f), (h), (i).
		(d)--(i) Results of the calculation on the ferromagnetic MTJ. 
		(d) TMR ratio $r_{\text{t}}$ (Eq.~(\ref{eq:TMR_ratio})) on the plane of $J$ and $E$.
		(e), (f) Transmissions as a function of the magnetic interaction, $J$, at (e) $E = -2$ and (f) $0$.
		(g) Ratio $r_{\text{c}}$ (Eq.~(\ref{eq:TMR_ratio_LDOS})).
		(h)--(i) Magnetic interaction, $J$, dependence of $g_{\text{c}}(E)$.
	}
	\label{fig:ferro_TMR}
\end{figure*}
%%%%%%%%%%%%%%%%%%%%%%%%%%%%%%%%%%%%%%%%%%%%%%%%%%%%
Let us first recall the conventional TMR, 
namely the TMR using the ferromagnetic electrodes as shown in Fig.~\ref{fig:MTJ_ferro_ferri}(a).
We set the localized spin moments as $\boldsymbol{s}_{i} = {}^{t}\begin{pmatrix} 0 & 0 & 1 \\ \end{pmatrix}$ for all sites in the left electrode.
For the sites in the right electrodes, 
we set $\boldsymbol{s}_{i} = {}^{t}\begin{pmatrix} 0 & 0 & 1 \\ \end{pmatrix}$ and ${}^{t}\begin{pmatrix} 0 & 0 & -1 \\ \end{pmatrix}$ for the parallel and antiparallel configurations, respectively.
The schematics of these two configurations are shown in Figs.~\ref{fig:ferro_TMR}(a) and \ref{fig:ferro_TMR}(b).
We set the hopping $t$ as a unit, $t = 1$.
The on-site potential is set as $\varepsilon_{i} = 0$ for the electrodes, 
and $\varepsilon_{i} = 10$ for the barrier region.
The system size is set as $L = 8$ and $W = 160$.
%%%%%%%%%%%%%%%%%%%%%%%%%%%%%%%%%%%%%%%%%%%%%%%%%%%%
\subsection{Bulk properties}
Before discussing the properties of the tunneling conductance, 
we see the properties of the bulk ferromagnetic metals used as the electrodes,
namely, the energy bands and the density of states (DOS).
The DOS is given as $\mathcal{N}_{\sigma}(E) = \int_{\text{BZ}} \mathrm{d}\boldsymbol{k} \ \delta(E - E_{\boldsymbol{k}, \sigma})$,
where $E_{\boldsymbol{k}, \sigma}$ is the energy band with spin-$\sigma$.
For the energy bands of the bulk electrodes,
we consider the two-dimensional square lattice described by $\mathcal{H}$ given in Eq.~(\ref{eq:hamiltonian_all}).
The energy bands are found to be $E_{\boldsymbol{k}}^{\pm} = -2t \left( \cos{k_{x}} + \cos{k_{y}}\right) \pm J$.
The DOS of the ferromagnet is shown in Fig.~\ref{fig:ferro_TMR}(c) at $J = 1$,
where the right-hand side and the left-hand side are the ones with the up-spin and down-spin, respectively.
By introducing a finite $J$, the energy band with up-spin gains the energy $-J$,
while that with down-spin shifts by $+J$.
%%%%%%%%%%%%%%%%%%%%%%%%%%%%%%%%%%%%%%%%%%%%%%%%%%%%
\subsection{Transmission and local density of states}
In Fig.~\ref{fig:ferro_TMR}(d), we show the TMR ratio defined as
\begin{align}
	r_{\text{t}} = \dfrac{T_{\text{P}} - T_{\text{AP}}}{T_{\text{P}} + T_{\text{AP}}},
	\label{eq:TMR_ratio}
\end{align}
on the plane of $J$ and $E$.
Here, $T_{\text{P/AP}}$ denotes the transmission for the parallel/antiparallel configuration.
We note that the definition above is slightly different from the conventional optimistic/pessimistic ones for the reason on normalization.
At $J = 0$, the whole system is nonmagnetic and has the degeneracy on the spin degrees of freedom, 
and thus $T_{\text{P}} = T_{\text{AP}}$ holds at each energy.
Namely, $r_{\text{t}}$ is zero.
When we introduce a finite magnetic interaction $J$, the degeneracy is lifted, 
and $T_{\text{P}}$ starts to take a larger value than $T_{\text{AP}}$;
a finite TMR ratio is observed.
Due to the asymmetric structure of the barrier in the energy,
$r_{\text{t}}$ is also asymmetric with respect to $E = 0$.
The $J$-dependence of $T_{\text{P}}$ and $T_{\text{AP}}$ is shown in Figs.~\ref{fig:ferro_TMR}(e) and \ref{fig:ferro_TMR}(f).
At $E = -2$, both of $T_{\text{P}}$ and $T_{\text{AP}}$ decrease when $J$ increases,
and $T_{\text{AP}}$ reaches zero at $J = 2$~\cite{accuracy}.
When $E = 0$, $T_{\text{P}}$ increases with $J$, while $T_{\text{AP}}$ decreases to zero.
\par
To better understand the transmission properties, we examine the LDOS in addition to the bulk DOS.
In the naive Julliere's picture, the product of the bulk DOS, $g(E)$, defined as
\begin{align}
	g(E) = \sum_{\sigma} \mathcal{N}_{\text{L}, \sigma}(E) \mathcal{N}_{\text{R}, \sigma}(E),
	\label{eq:dos_product}
\end{align}
describes the transmission~\cite{Julliere1975_PhysLettA_54A_225,Maekawa1982_IEEETranMagn_18_707},
where $\mathcal{N}_{\text{L/R}, \sigma}(E)$ is the bulk DOS with spin-$\sigma$, $\mathcal{N}_{\sigma}(E)$, of the left/right electrodes.
However, $g(E)$ does not consider the barrier, and thus this picture holds only for the limited cases.
Instead, the LDOS has been utilized to capture the details on the MTJs.
In particular, it has been proposed that the transmission can be described using the LDOS at the interfaces of the MTJ.
In fact, when the potential of the barrier is high enough,
the conductance derived by the Kubo formula~\cite{Mathon1997_PhysRevB_56_11810} is proportional to the product of the LDOS at the left and right interfaces, 
and to the exponential function, $\mathrm{e}^{-\kappa L}$, 
representing the decay inside the barrier~\cite{Tsymbal2005_JApplPhys_97_10C910,Tsymbal2007_ProgMaterSci_52_401}.
Here, $\kappa$ stands for the decaying property, namely, 
$1/\kappa$ means the spin diffusion length~\cite{Shim2008_PhysRevLett_100_226603}.
Hence, we consider the product of the LDOS at the interfaces, $g_{\text{i}}(E)$ given as
\begin{align}
	g_{\text{i}}(E)
	= \dfrac{1}{2} \sum_{\sigma}
	& \left[
	N_{\sigma} \left( E; 1, \dfrac{W}{2} \right) N_{\sigma}\left( E; L, \dfrac{W}{2} \right)
	\right. \notag \\
	& \hspace{-10mm}\left.
	+ N_{\sigma} \left( E; 1, \dfrac{W}{2}+1 \right) N_{\sigma}\left( E; L, \dfrac{W}{2}+1 \right)
	\right],
	\label{eq:ldos_product_interface}
\end{align}
where $N_{\sigma}(E; x, y)$ ($1 \leq x \leq L$, $1 \leq y \leq W$) is the LDOS of the barrier at $(x, y)$.
Since we impose the open boundary condition in the $y$-direction,
we take the average over the sites at $y = W/2$ and $W/2+1$ in Eqs.~(\ref{eq:ldos_product_interface}) and (\ref{eq:ldos_product_center}) to reduce the effects of the oscillation due to the boundary,
and we implicitly assume that the spins do not flip inside the barrier~\cite{evenodd}.
However, in general, it is not easy to precisely evaluate the exponent $\kappa$; we should need the transmission coefficients~\cite{Zhang1999_EurPhysJB_10_599} in addition to finding the electronic structures.
\par
Alternatively, we here utilize the LDOS at the center of the barrier.
Expecting that it contains both the details of the MTJ and the decay,
we consider the product as follows;
\begin{align}
	g_{\text{c}}(E)
	= \dfrac{1}{2} \sum_{\sigma}
	& \left[
	N_{\sigma} \left( E; \dfrac{L}{2}, \dfrac{W}{2} \right) N_{\sigma}\left( E; \dfrac{L}{2}+1, \dfrac{W}{2} \right)
	\right. \notag \\
	& \hspace{-10mm}\left.
	+ N_{\sigma} \left( E; \dfrac{L}{2}, \dfrac{W}{2}+1 \right) N_{\sigma}\left( E; \dfrac{L}{2}+1, \dfrac{W}{2}+1 \right)
	\right].
	\label{eq:ldos_product_center}
\end{align}
In this scheme, we only have to find the electronic structures to estimate the transmission.
While we now consider the case with even-$L$ and calculate the product of the LDOS at $x = L/2$ and $L/2+1$, 
we should replace both of them with the one at $x = (L+1)/2$ for the odd-$L$ case.
\par
Similarly to the TMR ratio $r_{\text{t}}$, we calculate the ratio, $r_{\text{c}}$, defined as
\begin{align}
	r_{\text{c}} = \dfrac{g_{\text{c}, \text{P}}(E) - g_{\text{c}, \text{AP}}(E)}{g_{\text{c}, \text{P}}(E) + g_{\text{c}, \text{AP}}(E)},
	\label{eq:TMR_ratio_LDOS}
\end{align}
where $g_{\text{c}, \text{P}}(E)$ and $g_{\text{c}, \text{AP}}(E)$ are $g_{\text{c}}(E)$ for the parallel and antiparallel configurations, respectively.
Figure~\ref{fig:ferro_TMR}(g) shows $r_{\text{c}}$ on the plane of $J$ and $E$.
We find that $r_{\text{c}}$ qualitatively reproduces the TMR ratio $r_{\text{t}}$ shown in Fig.~\ref{fig:ferro_TMR}(d).
We show the $J$-dependence of $g_{\text{c}}(E)$ in Figs.~\ref{fig:ferro_TMR}(h) and \ref{fig:ferro_TMR}(i),
and that of $g(E)$ in the insets.
At $E = -2$, $g_{\text{c}}(E)$ reproduces the overall properties of the transmission,
while $g(E)$ increases with $J$ and does not reproduce the transmission properties.
When we increase the energy to $E = 0$,
the bulk DOS with the up and down spins take the same values,
and the estimation in terms of $g(E)$ would lead to the absence of the TMR.
Even in this case, the product of the LDOS at the center of the barrier well reproduces the transmission properties;
the $J$-dependence of $g_{\text{c}}(E)$ is qualitatively the same as the one of the transmissions.
\par
We remark that we also examine the $J$-dependence of $g_{\text{i}}(E)$ following the previous proposals~\cite{Mathon1997_PhysRevB_56_11810,Tsymbal2005_JApplPhys_97_10C910,Tsymbal2007_ProgMaterSci_52_401}.
We find that $g_{\text{c}}(E)$ better reproduce the transmission properties than $g_{\text{i}}(E)$ when we compare these two quantities
(see Appendix~\ref{sec:appendix_ldos_interface} for the detailed results of $g_{\text{i}}(E)$),
which is due to the absence of the decay effect in $g_{\text{i}}(E)$.
%%%%%%%%%%%%%%%%%%%%%%%%%%%%%%%%%%%%%%%%%%%%%%%%%%%%
\section{Tunneling magnetoresistance effect with ferrimagnetic/antiferromagnetic electrodes}
\label{sec:ferrimagnetic_TMR}
%%%%%%%%%%%%%%%%%%%%%%%%%%%%%%%%%%%%%%%%%%%%%%%%%%%%
\begin{figure}[t]
	\centering
	\includegraphics[width=86mm]{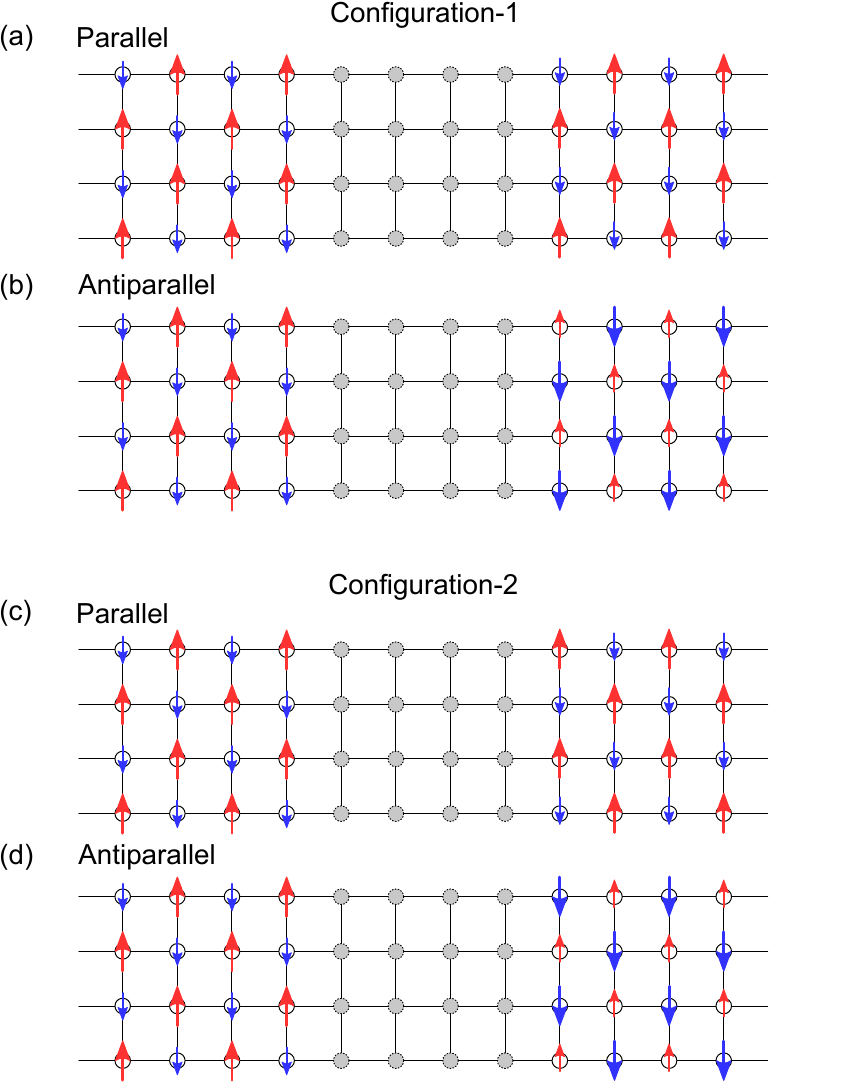}
	\caption{%
		Two configurations of the magnetic tunnel junctions (MTJ) using the ferrimagnetic electrodes, configuration-1 and 2. 
		(a) Parallel and (b) antiparallel alignments of configuration-1.
		(c) Parallel and (d) antiparallel alignments of configuration-2.
	}
	\label{fig:MTJ_ferri}
\end{figure}
%%%%%%%%%%%%%%%%%%%%%%%%%%%%%%%%%%%%%%%%%%%%%%%%%%%%
Next, we examine the tunneling conductance of the MTJ with ferrimagnetic electrodes, including the antiferromagnet, 
and apply the analysis by means of the LDOS.
We here concentrate on the ferrimagnet with two-sublattices, A and B, on the square lattice.
We assume that the ferrimagnet has the G-type structure;
all spins at the nearest-neighbors of the spins in the A-sublattice are in the B-sublattice, and vice versa.
In the ferrimagnetic MTJ, we define the parallel and antiparallel configurations by the alignments of the spins on the same sublattice.
Due to the two-sublattice structure, 
there are two pairs of the parallel--antiparallel configurations whose schematic views are shown in Figs.~\ref{fig:MTJ_ferri}(a)--\ref{fig:MTJ_ferri}(d).
In the first one, which we call configuration-1,
two localized spins next to the barrier layer on the left and right electrodes with the same $y$-coordinates are in the different sublattices.
The parallel and antiparallel arrangements of configuration-1 is shown in Figs.~\ref{fig:MTJ_ferri}(a) and \ref{fig:MTJ_ferri}(b), respectively.
In the second one, which we refer to as configuration-2, those two spins are in the same sublattices, 
whose parallel and antiparallel arrangements are respectively shown in Figs.~\ref{fig:MTJ_ferri}(c) and \ref{fig:MTJ_ferri}(d).
The parameters in the Hamiltonian are taken in common with the ferromagnetic MTJ;
$\varepsilon_{i} = 0$ and $10t$ for the electrodes and the barrier respectively, 
and $L = 8$ and $W = 160$.
%%%%%%%%%%%%%%%%%%%%%%%%%%%%%%%%%%%%%%%%%%%%%%%%%%%%
\subsection{Bulk properties}
\begin{figure}[t]
	\centering
	\includegraphics[width=86mm]{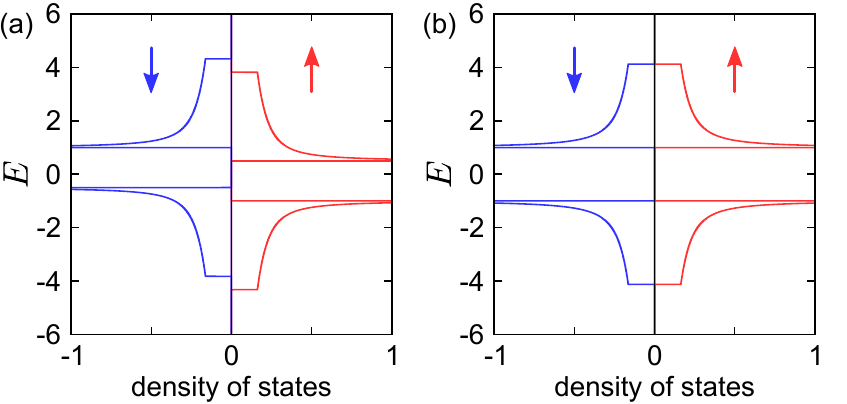}
	\caption{%
		Spin-resolved density of states of (a) the two-dimensional ferrimagnet with $(s_{\text{A}}, s_{\text{B}}) = (1.0, -0.5)$,
		and (b) the antiferromagnet with $(s_{\text{A}}, s_{\text{B}}) = (1.0, -1.0)$.
		For both cases, $J = 1.0$.
	}
	\label{fig:ferri_antiferro_DOS}
\end{figure}

We first see the bulk properties of the ferrimagnetic electrodes;
we calculate the energy band and the DOS of the system described by the Hamiltonian $\mathcal{H}$ on the square lattice.
The spins of A- and B-sublattices are set as $\boldsymbol{s}_{\text{A}} = {}^{t}\begin{pmatrix} 0 & 0 & s_{\text{A}} \\ \end{pmatrix}$ and $\boldsymbol{s}_{\text{B}} = {}^{t}\begin{pmatrix} 0 & 0 & s_{\text{B}} \\ \end{pmatrix}$,
respectively.
Since the ferrimagnet has two-sublattices,
there are two energy bands, $E_{\boldsymbol{k}, \sigma}^{\pm}$, for each spin degrees of freedom $\sigma$.
The energy bands are written as
\begin{align}
	E_{\boldsymbol{k}, \uparrow}^{\pm}
& = \dfrac{- J \left( s_{\mathrm{A}} + s_{\mathrm{B}} \right)
						\pm \sqrt{J^{2} \left( s_{\mathrm{A}} + s_{\mathrm{B}} \right)^{2}
						- 4 \left( J^{2} s_{\mathrm{A}} s_{\mathrm{B}} - \gamma_{\boldsymbol{k}}^{2} \right)}
					}
					{2}, \label{eq:ferri_energyband_up} \\
	E_{\boldsymbol{k}, \downarrow}^{\pm}
& = \dfrac{+ J \left( s_{\mathrm{A}} + s_{\mathrm{B}} \right)
						\pm \sqrt{J^{2} \left( s_{\mathrm{A}} + s_{\mathrm{B}} \right)^{2}
					- 4 \left( J^{2} s_{\mathrm{A}} s_{\mathrm{B}} - \gamma_{\boldsymbol{k}}^{2} \right)}
					}
					{2}, \label{eq:ferri_energyband_dn}
\end{align}
where $\gamma_{\boldsymbol{k}} = - 2t \left( \cos{k_{x}} + \cos{k_{y}} \right)$.
In Figs.~\ref{fig:ferri_antiferro_DOS}(a) and \ref{fig:ferri_antiferro_DOS}(b), 
we show examples of the DOS of the ferrimagnet with $(s_{\text{A}}, s_{\text{B}}) = (1.0, -0.5)$, 
and that of the antiferromagnet with $(s_{\text{A}}, s_{\text{B}}) = (1.0, -1.0)$, respectively.
%%%%%%%%%%%%%%%%%%%%%%%%%%%%%%%%%%%%%%%%%%%%%%%%%%%%
%%%%%%%%%%%%%%%%%%%%%%%%%%%%%%%%%%%%%%%%%%%%%%%%%%%%
\subsection{Transmissions and local density of states}
\subsubsection{Ferrimagnetic electrode}
\label{subsubsec:ferrimagnetic_TMR}
\begin{figure*}[t]
	\centering
	\includegraphics[width=172mm]{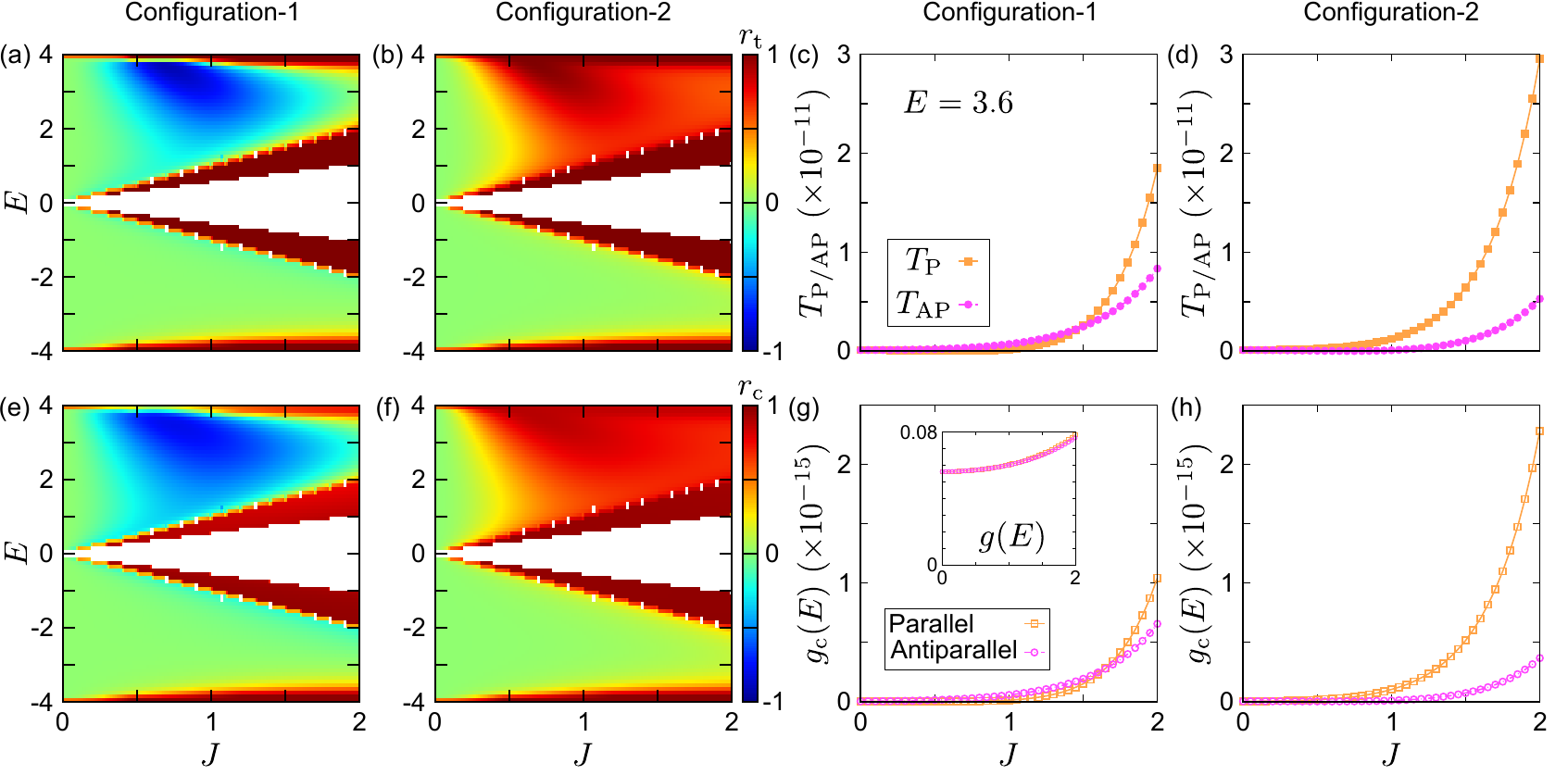}
	\caption{%
		Results of the transmission calculation for the ferrimagnetic tunnel junction with $(s_{\text{A}}, s_{\text{B}}) = (1.0, -0.5)$ for configuration-1 and 2.
		(a), (b) TMR ratio $r_{\text{t}}$ (Eq.~(\ref{eq:TMR_ratio})) on the plane of $J$ and $E$.
		(c), (d) Transmissions at $E = 3.6$ with respect to $J$.
		(e), (f) Ratio $r_{\text{c}}$ (Eq.~(\ref{eq:TMR_ratio_LDOS})).
		(g), (h) Product of the local density of states at the center of the barrier, $g_{\text{c}}(E)$, as a function of $J$.
		Inset in (g) is the $J$-dependence of $g(E)$.
	}
	\label{fig:ferri_TMR_center}
\end{figure*}
%%%%%%%%%%%%%%%%%%%%%%%%%%%%%%%%%%%%%%%%%%%%%%%%%%%%
We discuss the transmission properties of the ferrimagnetic MTJ.
First, we investigate the system where we fix the magnetization of two sublattices as $(s_{\text{A}}, s_{\text{B}}) = (1.0, -0.5)$.
Figures~\ref{fig:ferri_TMR_center}(a) and \ref{fig:ferri_TMR_center}(b) show the TMR ratio, $r_{\text{t}}$, on the plane of $J$ and $E$ for configuration-1 and 2, respectively.
The $J$-dependence of the transmissions for configuration-1 and 2 at $E = 3.6$ is respectively shown in Figs.~\ref{fig:ferri_TMR_center}(c) and \ref{fig:ferri_TMR_center}(d).
Without the $s$--$d$ coupling $J$, the localized magnetic moment does not affect the transmission, 
and $T_{\text{P}}$ and $T_{\text{AP}}$ take the same values for both configuration-1 and 2.
When a small $J$ is introduced in configuration-1, $T_{\text{AP}}$ is larger than $T_{\text{P}}$, 
and $T_{\text{P}}$ becomes larger than $T_{\text{AP}}$ at $J \simeq 1.5$.
Meanwhile, $T_{\text{P}}$ is larger than $T_{\text{AP}}$ at finite $J$ in configuration-2.
\par
As well as the ferromagnetic MTJ discussed in Sec.~\ref{sec:ferrimagnetic_TMR},
we focus on $g_{\text{c}}(E)$ defined by Eq.~(\ref{eq:ldos_product_center}) and calculate $r_{\text{c}}$ given in Eq.~(\ref{eq:TMR_ratio_LDOS}).
In Figs.~\ref{fig:ferri_TMR_center}(e) and \ref{fig:ferri_TMR_center}(f), 
we plot $r_{\text{c}}$ for configuration-1 and 2, respectively, 
which indicates that $g_{\text{c}}(E)$ qualitatively traces the TMR property also in the ferrimagnetic MTJs.
We plot the $J$-dependence of $g_{\text{c}}(E)$ in Figs.~\ref{fig:ferri_TMR_center}(g) and \ref{fig:ferri_TMR_center}(h) for configuration-1 and 2 at $E = 3.6$, respectively,
together with $g(E)$ given in Eq.~(\ref{eq:dos_product}) in the inset of Fig.~\ref{fig:ferri_TMR_center}(g).
We can see that the transmission and $g_{\text{c}}(E)$ similarly changes with $J$,
while $g(E)$ does not reproduce the transmission.
Here the average is meaningful also for capturing the two-sublattice magnetic structure in the $y$-direction in the ferrimagnetic MTJ, 
whereas in the ferromagnetic MTJ the average over $y = W/2$ and $W/2+1$ is important to take the open boundary conditions into account.
We note that $g_{\text{c}}(E)$ traces even the reversal of the transmission occurring in configuration-1,
whereas $g(E)$ or $g_{\text{i}}(E)$ defined by the interfacial DOS does not (see Fig.~\ref{fig:appendix_ferri_TMR_interface} for details).
%%%%%%%%%%%%%%%%%%%%%%%%%%%%%%%%%%%%%%%%%%%%%%%%%%%%
\subsubsection{Antiferromagnetic electrode}
\label{subsubsec:antiferromagnetic_TMR}
%%%%%%%%%%%%%%%%%%%%%%%%%%%%%%%%%%%%%%%%%%%%%%%%%%%%
\begin{figure}[t]
	\centering
	\includegraphics[width=86mm]{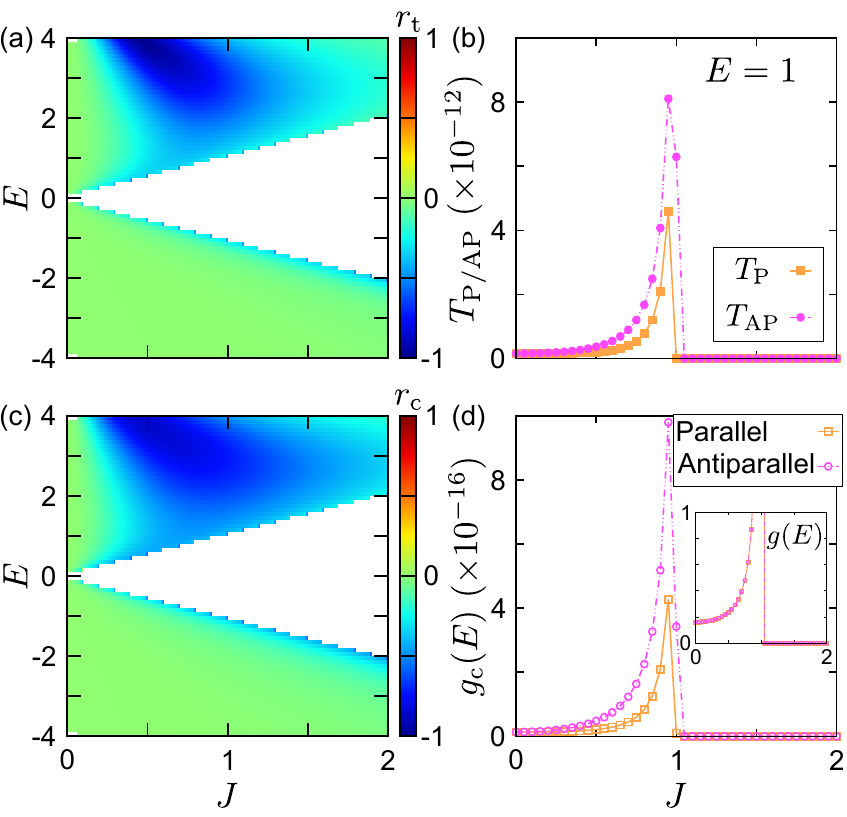}
	\caption{%
		Results of the transmission calculation of the antiferromagnetic tunnel junction with $(s_{\text{A}}, s_{\text{B}}) = (1.0, -1.0)$. 
		(a) TMR ratio $r_{\text{t}}$ (Eq.~(\ref{eq:TMR_ratio})) on the plane of $J$ and $E$.
		(b) Transmissions for the parallel and antiparallel configurations at $E = 1$.
		(c) Ratio $r_{\text{c}}$ (Eq.~(\ref{eq:TMR_ratio_LDOS})).
		(d) Magnetic coupling, $J$, dependence of $g_{\text{c}}(E)$ at $E = 1$.
		Inset is the $J$-dependence of $g(E)$.
	}
	\label{fig:antiferro_TMR_center}
\end{figure}
%%%%%%%%%%%%%%%%%%%%%%%%%%%%%%%%%%%%%%%%%%%%%%%%%%%
Next we discuss the antiferromagnetic limit, $(s_{\text{A}}, s_{\text{B}}) = (1.0, -1.0)$.
In this case, configuration-1 and 2 are degenerate; 
the parallel and antiparallel alignments of configuration-1 correspond to the antiparallel and parallel alignments of configuration-2, respectively.
Hence, here we consider configuration-1 only and define the parallel and antiparallel configurations by configuration-1.
We show the TMR ratio $r_{\text{t}}$ (Eq.~(\ref{eq:TMR_ratio})) in Fig.~\ref{fig:antiferro_TMR_center}(a),
and the $J$-dependence of the transmission at in Fig~\ref{fig:antiferro_TMR_center}(b) at $E = 1$.
At $E = 1$, both of $T_{\text{P}}$ and $T_{\text{AP}}$ increase with $J$ at $J \lesssim 1$ as shown in Fig.~\ref{fig:antiferro_TMR_center}(b).
At $J \gtrsim 1$, $\mathcal{N}_{\sigma}(E)$ is zero,
namely, there is no incidence from the electrodes.
Thus, the transmission of each configuration becomes zero.
\par
Figures~\ref{fig:antiferro_TMR_center}(c) represents the ratio $r_{\text{c}}$ (Eq.~(\ref{eq:TMR_ratio_LDOS})).
We confirm that $r_{\text{c}}$ has a parameter dependence qualitatively the same as the one of $r_{\text{t}}$.
In fact, the $J$-dependence of $g_{\text{c}}(E)$ at $E = 1$ shown in Fig.~\ref{fig:antiferro_TMR_center}(d) well reproduce the $J$-dependence of the transmissions shown in Fig.~\ref{fig:antiferro_TMR_center}(b).
In the inset of Fig.~\ref{fig:antiferro_TMR_center}(d) we show $g(E)$ for the parallel and antiparallel configurations, which are degenerate and do not predict a finite TMR effect.
We note that $g_{\text{i}}(E)$ defined by the interfacial LDOS has a parameter dependence qualitatively different from the one of the transmissions (see Fig.~\ref{fig:appendix_antiferro_TMR_interface} for details).
%%%%%%%%%%%%%%%%%%%%%%%%%%%%%%%%%%%%%%%%%%%%%%%%%%%%
\section{Discussion}
\label{sec:remarks}
%%%%%%%%%%%%%%%%%%%%%%%%%%%%%%%%%%%%%%%%%%%%%%%%%%%%
\subsection{Hierarchy in estimating the transmission properties}
%%%%%%%%%%%%%%%%%%%%%%%%%%%%%%%%%%%%%%%%%%%%%%%%%%%%
\begin{figure}[t]
	\centering
	\includegraphics[width=86mm]{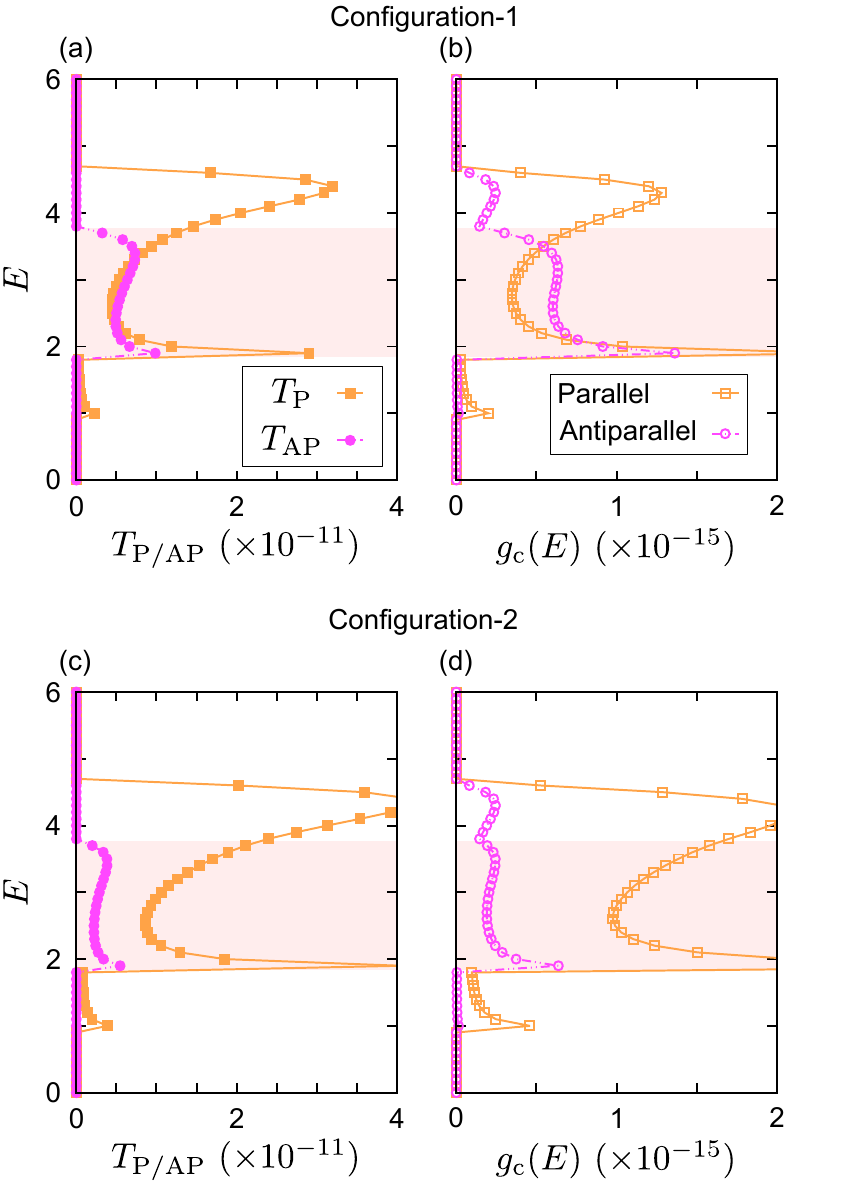}
	\caption{%
		Results of the transmission calculations for the ferrimagnetic tunnel junction with $(s_{\text{A}}, s_{\text{B}}) = (1.0, -0.5)$ at $J = 1.85$ for (a), (b) configuration-1 and (c), (d) 2.
		(a), (c) Transmissions and (b), (d) $g_{\text{c}}(E)$ for the parallel and antiparallel alignments.
		Shaded regions represent the energies where both spin-states have finite DOS (see Fig.~\ref{fig:ferri_antiferro_DOS}(a)).
	}
	\label{fig:ferri_TMR_Edep}
\end{figure}
%%%%%%%%%%%%%%%%%%%%%%%%%%%%%%%%%%%%%%%%%%%%%%%%%%%%
We presented a hierarchy in the evaluation of the transmission properties in Fig.~\ref{fig:concept}.
Here we discuss how the appearance, namely, the bulk DOS and the spin configurations,
and the LDOS are related to the transmission properties.
For the ferrimagnetic tunnel junction with $(s_{\text{A}}, s_{\text{B}}) = (1.0, -0.5)$ at $J = 1.85$,
we show the energy dependence of the transmissions and $g_{\text{c}}(E)$ in Fig.~\ref{fig:ferri_TMR_Edep}.
Figures~\ref{fig:ferri_TMR_Edep}(a) and \ref{fig:ferri_TMR_Edep}(b) are the results for configuration-1,
and Figs.~\ref{fig:ferri_TMR_Edep}(c) and \ref{fig:ferri_TMR_Edep}(d) are for configuration-2.
When either of the two spin-states has a finite DOS, 
$T_{\text{P}}$ takes a larger value than $T_{\text{AP}}$,
which supports the Julliere's picture.
\par
When both spin-states have finite DOS, shown as the shaded regions in Fig.~\ref{fig:ferri_TMR_Edep},
$T_{\text{AP}}$ is sometimes larger than $T_{\text{P}}$,
e.g., at $E \sim 3$ for configuration-1 (Fig.~\ref{fig:ferri_TMR_Edep}(a)).
This means that the conventional Julliere's description breaks down in these regions.
Instead, the spin configurations at the interface usually gives very rough estimation of the transmission.
Let us focus on the spin configurations at the interface.
As shown in Figs.~\ref{fig:MTJ_ferri}(a) and \ref{fig:MTJ_ferri}(b), 
for configuration-1, the interfacial spins of the two electrodes with the same $y$-coordinates align antiparallelly in the parallel arrangement, 
and those spins in the antiparallel arrangement align parallelly.
For the configuration-2, the parallel arrangement have the interfacial spins with opposite directions,
and the antiparallel arrangement have the interfacial spins pointing the same directions,
which is represented in Figs.~\ref{fig:MTJ_ferri}(c) and \ref{fig:MTJ_ferri}(d).
Since the spins unlikely to flip through coherent tunneling, 
the transmissions become larger when the interfacial spins of two electrodes align parallelly.
In fact, in the case where two spin-states have nonzero DOS, 
$T_{\text{P}} < T_{\text{AP}}$ basically holds for configuration-1,
whereas $T_{\text{P}} > T_{\text{AP}}$ holds in a broad $E$-region for configuration-2 (see Figs.~\ref{fig:ferri_TMR_Edep}(a) and \ref{fig:ferri_TMR_Edep}(c)).
\par
From the interfacial spin configurations, we can roughly predict the TMR properties in many cases.
As shown in Fig.~\ref{fig:ferri_TMR_Edep}(a), however,
$T_{\text{P}}$ becomes larger than $T_{\text{AP}}$ contrary to the prediction from the interfacial spins at $E \sim 3.7$ in configuration-1 with the bulk DOS consisting of both spin-states.
Still in this case, $g_{\text{c}}(E)$ for the parallel configuration takes a larger value than that for the antiparallel configuration.
Furthermore, $g_{\text{c}}(E)$ gives us the detailed information on the parameter dependence as shown in Figs.~\ref{fig:ferri_TMR_Edep}(b) and \ref{fig:ferri_TMR_Edep}(d), 
while we can only know from the appearance whether the parallel or antiparallel configuration gives the larger transmission.
To completely understand the transmission properties, of course we should calculate transmission itself,
but we expect that the estimation in terms of the LDOS is enough as an initial way.
%%%%%%%%%%%%%%%%%%%%%%%%%%%%%%%%%%%%%%%%%%%%%%%%%%%%
\subsection{Details of the magnetic tunnel junctions}
We have considered the simplest cases,
where the electronic orbitals are isotropic and the barrier is structureless.
In reality, we should consider the details of the MTJs.
In the Fe(001)/MgO(001)/Fe epitaxial MTJ, for example, 
the $\Delta_{1}$-symmetry state with a large spin polarization has less decay,
which dominantly contributes to the large TMR ratio~\cite{Butler2001_PhysRevB_63_054416,Mathon2001_PhysRevB_63_220403}.
If we focus on this $\Delta_{1}$ Bloch state,
we can apply our treatment and estimate the TMR effect in terms of the LDOS.
Besides, due to structural disorders or hybridization of orbitals,
the electronic and magnetic states at the interfaces may be modulated.
We can trace the effect of the modulation with $g_{\text{c}}(E)$,
which is indicated by the results that $g_{\text{c}}(E)$ reproduces the transmissions with each of two different interfaces, configuration-1 and 2, for the ferrimagnetic MTJs.
\par
In the typical metals used in the ferrimagnetic spintronics such as $\mathrm{GdFeCo}$~\cite{Stanciu2006_PhysRevB_73_220402,Stanciu2007_PhysRevLett_99_217204,Radu2011_Nature_472_205} or $\mathrm{Mn_{2}Ru}_{x}\mathrm{Ga}$~\cite{Kurt2014_PhysRevLett_112_027201},
the valences of the ions carrying the magnetic moments of the different sublattices are different.
This charge inequivalence determines the interfacial structure to keep the charge neutrality at the interface.
Namely, if we also take the charge degrees of freedom into account, 
we can in principle design the interfacial magnetic structure.
On the other hand, when we cannot control the interfacial structures precisely,
the averaged structure of configuration-1 and 2 seems to be realized,
which can be regarded as the ferromagnetic MTJ of the net magnetic moments.
We have numerically confirmed that the Julliere's picture with the bulk DOS holds like ferromagnets in that case.
\par
For antiferromagnetic MTJs, when we use the antiferromagnets with the macroscopic time-reversal symmetry, configuration-1 and 2 are not distinguished.
The TMR effect then vanishes since there is a degeneracy on the transmission between configuration-1 and 2 as mentioned in Sec.~\ref{sec:ferrimagnetic_TMR}.
By contrast, the antiferromagnets macroscopically breaking the time-reversal symmetry separate the MTJs with configuration-1 and 2,
which enables us to observe a finite TMR effect in the antiferromagnetic MTJs.
Actually, the MTJs using such antiferromagnets have been theoretically proposed~\cite{Shao2021_NatCommun_12_7061,Smejkal2022_PhysRevX_12_011028,Dong2022_PhysRevLett_128_197201}.
%%%%%%%%%%%%%%%%%%%%%%%%%%%%%%%%%%%%%%%%%%%%%%%%%%%%
\section{Summary and perspectives}
\label{sec:summary}
In summary, we have numerically studied the tunneling magnetoresistance (TMR) effect modelizing the magnetic tunnel junction (MTJ) consisting of the ferrimagnetic electrodes as well as the well-known ferromagnetic ones.
To grasp the transmission properties, we have focused on the local density of states.
We have shown that the transmission properties can be qualitatively reproduced by the product of the local density of states at the center of the barrier.
In the physical aspect, there can be multiple configurations for the ferrimagnetic MTJs owing to the sublattice structure of the electrodes.
Those multiple configurations give the different transmission properties,
and thus we should be careful for the magnetic configurations in the ferrimagnetic TMR.
\par
Our approach can be applied to more complicated cases where the detailed structures are taken into account.
When one performs a more realistic TMR calculation, 
the electronic structures and the wave-functions should be obtained from first-principles.
To calculate the transmission by using first-principles wave-functions, 
the methods such as the nonequilibrium Green's function formalism ~\cite{Taylor2001_PhysRevB_63_245407,Taylor2001_PhysRevB_63_121104,Brandbyge2002_PhysRevB_65_165401} or the scattering problem approach~\cite{Choi1999_PhysRevB_59_2267,Smogunov2004_PhysRevB_70_045417,DalCorso2005_PhysRevB_71_115106,DalCorso2006_PhysRevB_74_045429} are widely adopted.
However, these methods usually demand huge numerical costs,
which probably has prevented us from exploring the MTJ using various materials.
The calculation of the local density of states is much less costly,
so that the approach with the local density of states will serve an easy means to search for the MTJ with high efficiency.
%%%%%%%%%%%%%%%%%%%%%%%%%%%%%%%%%%%%%%%%%%%%%%%%%%%%
\begin{acknowledgments}
This work was supported by JST-Mirai Program (JPMJMI20A1), a Grant-in-Aid for Scientific Research (No. 21H04437, No. 21H04990, and No. 19H05825), and JST-PRESTO (JPMJPR20L7).
\end{acknowledgments}
%%%%%%%%%%%%%%%%%%%%%%%%%%%%%%%%%%%%%%%%%%%%%%%%%%%%
\appendix*
%%%%%%%%%%%%%%%%%%%%%%%%%%%%%%%%%%%%%%%%%%%%%%%%%%%%
\section{Product of the local density of states at the interfaces}
\label{sec:appendix_ldos_interface}
%%%%%%%%%%%%%%%%%%%%%%%%%%%%%%%%%%%%%%%%%%%%%%%%%%%%
\begin{figure}[t]
	\centering
	\includegraphics[width=86mm]{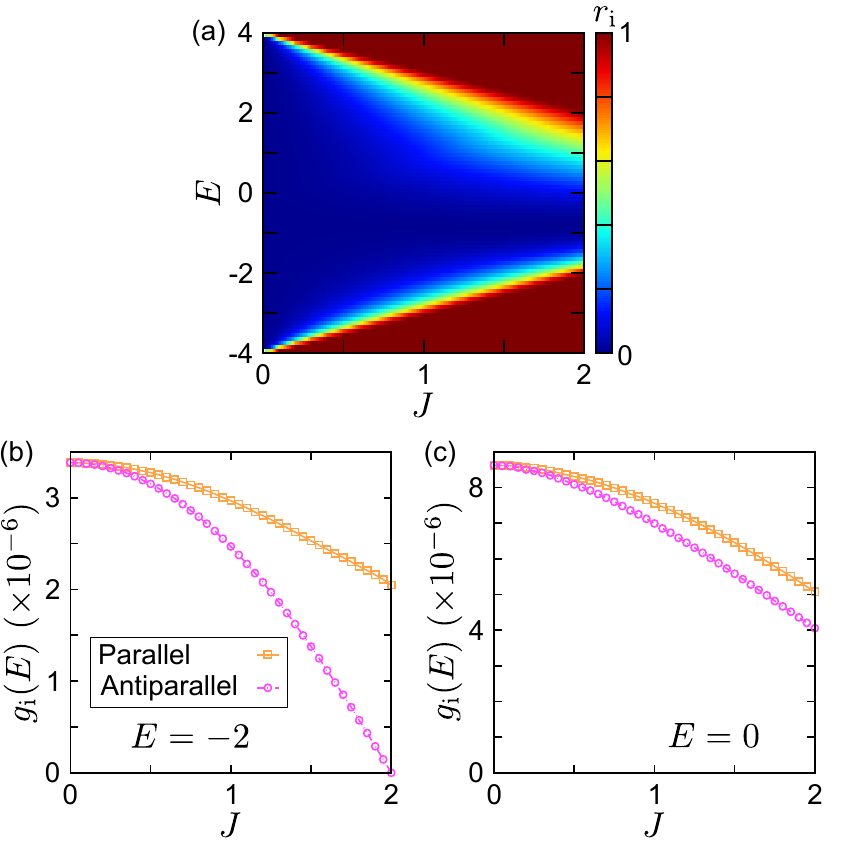}
	\caption{%
		Results of the simulation for ferromagnetic tunnel junctions (Sec.~\ref{sec:ferromagnetic_TMR}).
		See also Fig.~\ref{fig:ferro_TMR} as a comparison.
		(a) Ratio $r_{\text{i}}$ (Eq.~(\ref{eq:TMR_ratio_LDOS_interface})) on the plane of $J$ and $E$.
		(b), (c) Product of the interfacial local density of states $g_{\text{i}}(E)$ (Eq.~(\ref{eq:ldos_product_interface})) with respect to $J$ at (b) $E = -2$ and (c) $0$.
	}
	\label{fig:appendix_ferro_TMR_interface}
\end{figure}
%%%%%%%%%%%%%%%%%%%%%%%%%%%%%%%%%%%%%%%%%%%%%%%%%%%%
\begin{figure}[t]
	\centering
	\includegraphics[width=86mm]{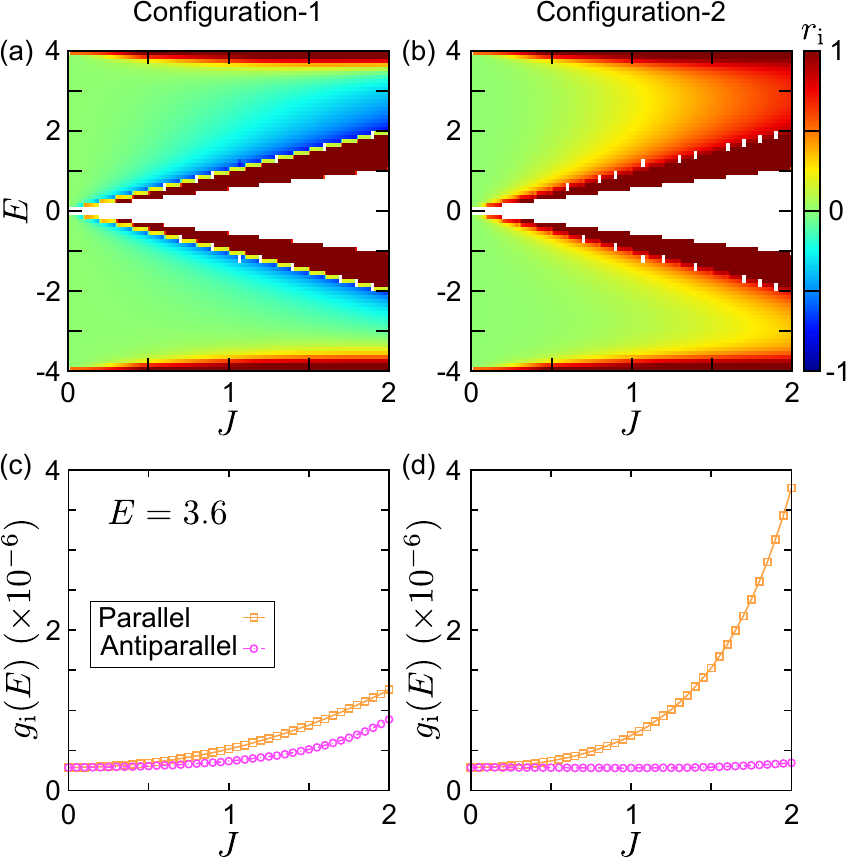}
	\caption{%
		Results of the calculation for ferrimagnetic tunnel junctions with $(s_{\text{A}}, s_{\text{B}}) = (1.0, -0.5)$ (Sec.~\ref{subsubsec:ferrimagnetic_TMR}).
		See also Fig.~\ref{fig:ferri_TMR_center} as a comparison.
		(a), (b) Ratio $r_{\text{i}}$ (Eq.~(\ref{eq:TMR_ratio_LDOS_interface})) on the plane of $J$ and $E$ for (a) configuration-1 and (b) 2.
		(c), (d) Product of the interfacial local density of states $g_{\text{i}}(E)$ (Eq.~(\ref{eq:ldos_product_interface})) as a function of $J$ at $E = 3.6$ for (c) configuration-1 and (d) 2.
	}
	\label{fig:appendix_ferri_TMR_interface}
\end{figure}
%%%%%%%%%%%%%%%%%%%%%%%%%%%%%%%%%%%%%%%%%%%%%%%%%%%%
\begin{figure}[t]
	\centering
	\includegraphics[width=86mm]{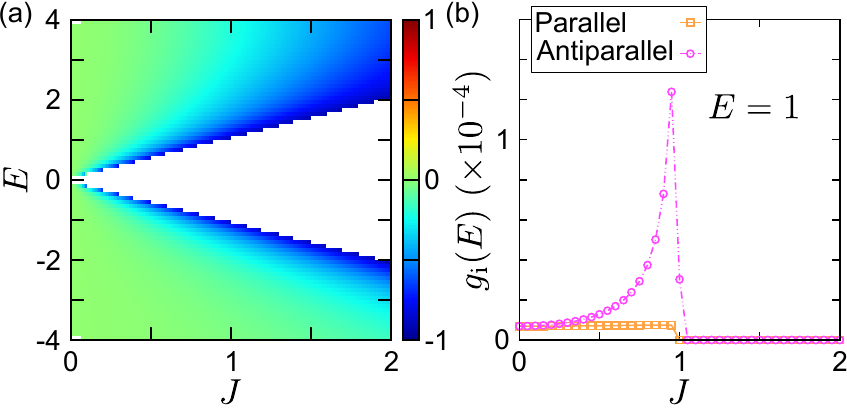}
	\caption{%
		Results of the simulation for antiferromagnetic tunnel junctions with $(s_{\text{A}}, s_{\text{B}}) = (1.0, -1.0)$ (Sec.~\ref{subsubsec:antiferromagnetic_TMR}).
		See also Fig.~\ref{fig:antiferro_TMR_center} as a comparison.
		(a) Ratio $r_{\text{i}}$ (Eq.~(\ref{eq:TMR_ratio_LDOS_interface})) on the plane of $J$ and $E$.
		(b) Product of the interfacial local density of states $g_{\text{i}}(E)$ (Eq.~(\ref{eq:ldos_product_interface})) at $E = 1$ as a function of $J$.
	}
	\label{fig:appendix_antiferro_TMR_interface}
\end{figure}
%%%%%%%%%%%%%%%%%%%%%%%%%%%%%%%%%%%%%%%%%%%%%%%%%%%%
In the main text, we have discussed the similarities between the transmissions and the product of the local density of states (LDOS) at the center of the barrier region.
Here we discuss the $J$-dependence of the product of the LDOS at the interface~\cite{Tsymbal2005_JApplPhys_97_10C910,Tsymbal2007_ProgMaterSci_52_401}.
Similarly to $r_{\text{c}}$ defined in Eq.~(\ref{eq:TMR_ratio_LDOS}), we define the ratio for the interfacial LDOS, $r_{\text{i}}$, as
\begin{align}
	r_{\text{i}} = \dfrac{g_{\text{i}, \text{P}}(E) - g_{\text{i}, \text{AP}}(E)}{g_{\text{i}, \text{P}}(E) + g_{\text{i}, \text{AP}}(E)}.
	\label{eq:TMR_ratio_LDOS_interface}
\end{align}
Here $g_{\text{i}, \text{P/AP}}$ is $g_{\text{i}}(E)$ for the parallel/antiparallel configuration.
We plot $r_{\text{i}}$ in Fig.~\ref{fig:appendix_ferro_TMR_interface}(a).
We find that $r_{\text{i}}$ also describes the transmission in the large- or small-$E$ regions where $r_{\text{t}}\simeq 1$ owing to the absence of the bulk DOS of either of the spin-states.
However, $r_{\text{i}}$ does not reproduce the intermediate-$E$ region.
Actually, as shown in Fig.~\ref{fig:appendix_ferro_TMR_interface}(b), 
$g_{\text{i}}(E)$ changes similarly to the transmission at $E = -2$ (see Fig.~\ref{fig:ferro_TMR}(e)),
whereas the $J$-dependence of $g_{\text{i}}(E)$ shown in Fig.~\ref{fig:appendix_ferro_TMR_interface}(c) largely deviates from that of the transmission at $E = 0$ (Fig.~\ref{fig:ferro_TMR}(f)).
\par
In Figs.~\ref{fig:appendix_ferri_TMR_interface}(a) and \ref{fig:appendix_ferri_TMR_interface}(b), 
we present $r_{\text{i}}$ for the ferrimagnetic MTJ with $(s_{\text{A}}, s_{\text{B}}) = (1.0, -0.5)$ for configuration-1 and 2, respectively.
Figures~\ref{fig:appendix_ferri_TMR_interface}(c) and \ref{fig:appendix_ferri_TMR_interface}(d) are the $J$-dependence of $g_{\text{i}}(E)$ at $E = 3.6$. 
These results shows that the transmission and $g_{\text{i}}(E)$ of configuration-2 may roughly agree with each other,
whereas the reversal of the transmission in configuration-1 is not reproduced by $g_{\text{i}}(E)$ at $J \leq 2$.
\par
Figure~\ref{fig:appendix_antiferro_TMR_interface}(a) shows $r_{\text{i}}$ for the antiferromagnetic MTJ; $(s_{\text{A}}, s_{\text{B}}) = (1.0, -1.0)$,
and Fig.~\ref{fig:appendix_antiferro_TMR_interface}(b) the $J$-dependence of $g_{\text{i}}(E)$ at $E = 1$.
At $E = 1$, $g_{\text{i}}(E)$ of the antiparallel configuration changes similarly to $T_{\text{AP}}$ as $J$ increases,
but the increase in $T_{\text{P}}$ at $J \lesssim 1$ is not observed in the $J$-dependence of $g_{\text{i}}(E)$ of the parallel configuration.
\par
Therefore, $g_{\text{i}}(E)$ is not enough to describe the transmission properties,
and the effect of decay inside the barrier should be additionally considered.


\begin{thebibliography}{74}%
\makeatletter
\providecommand \@ifxundefined [1]{%
 \@ifx{#1\undefined}
}%
\providecommand \@ifnum [1]{%
 \ifnum #1\expandafter \@firstoftwo
 \else \expandafter \@secondoftwo
 \fi
}%
\providecommand \@ifx [1]{%
 \ifx #1\expandafter \@firstoftwo
 \else \expandafter \@secondoftwo
 \fi
}%
\providecommand \natexlab [1]{#1}%
\providecommand \enquote  [1]{``#1''}%
\providecommand \bibnamefont  [1]{#1}%
\providecommand \bibfnamefont [1]{#1}%
\providecommand \citenamefont [1]{#1}%
\providecommand \href@noop [0]{\@secondoftwo}%
\providecommand \href [0]{\begingroup \@sanitize@url \@href}%
\providecommand \@href[1]{\@@startlink{#1}\@@href}%
\providecommand \@@href[1]{\endgroup#1\@@endlink}%
\providecommand \@sanitize@url [0]{\catcode `\\12\catcode `\$12\catcode
  `\&12\catcode `\#12\catcode `\^12\catcode `\_12\catcode `\%12\relax}%
\providecommand \@@startlink[1]{}%
\providecommand \@@endlink[0]{}%
\providecommand \url  [0]{\begingroup\@sanitize@url \@url }%
\providecommand \@url [1]{\endgroup\@href {#1}{\urlprefix }}%
\providecommand \urlprefix  [0]{URL }%
\providecommand \Eprint [0]{\href }%
\providecommand \doibase [0]{http://dx.doi.org/}%
\providecommand \selectlanguage [0]{\@gobble}%
\providecommand \bibinfo  [0]{\@secondoftwo}%
\providecommand \bibfield  [0]{\@secondoftwo}%
\providecommand \translation [1]{[#1]}%
\providecommand \BibitemOpen [0]{}%
\providecommand \bibitemStop [0]{}%
\providecommand \bibitemNoStop [0]{.\EOS\space}%
\providecommand \EOS [0]{\spacefactor3000\relax}%
\providecommand \BibitemShut  [1]{\csname bibitem#1\endcsname}%
\let\auto@bib@innerbib\@empty
%</preamble>
\bibitem [{\citenamefont {\ifmmode \check{Z}\else
  \v{Z}\fi{}uti\ifmmode~\acute{c}\else \'{c}\fi{}}\ \emph
  {et~al.}(2004)\citenamefont {\ifmmode \check{Z}\else
  \v{Z}\fi{}uti\ifmmode~\acute{c}\else \'{c}\fi{}}, \citenamefont {Fabian},\
  and\ \citenamefont {Das~Sarma}}]{Zutic2004_RevModPhys_76_323}%
  \BibitemOpen
  \bibfield  {author} {\bibinfo {author} {\bibfnamefont {I.}~\bibnamefont
  {\ifmmode \check{Z}\else \v{Z}\fi{}uti\ifmmode~\acute{c}\else \'{c}\fi{}}},
  \bibinfo {author} {\bibfnamefont {J.}~\bibnamefont {Fabian}}, \ and\ \bibinfo
  {author} {\bibfnamefont {S.}~\bibnamefont {Das~Sarma}},\ }\href {\doibase
  10.1103/RevModPhys.76.323} {\bibfield  {journal} {\bibinfo  {journal} {Rev.
  Mod. Phys.}\ }\textbf {\bibinfo {volume} {76}},\ \bibinfo {pages} {323}
  (\bibinfo {year} {2004})}\BibitemShut {NoStop}%
\bibitem [{\citenamefont {Felser}\ \emph {et~al.}(2007)\citenamefont {Felser},
  \citenamefont {Fecher},\ and\ \citenamefont
  {Balke}}]{Felser2007_AngewChemIntEd_46_668}%
  \BibitemOpen
  \bibfield  {author} {\bibinfo {author} {\bibfnamefont {C.}~\bibnamefont
  {Felser}}, \bibinfo {author} {\bibfnamefont {G.}~\bibnamefont {Fecher}}, \
  and\ \bibinfo {author} {\bibfnamefont {B.}~\bibnamefont {Balke}},\ }\href
  {\doibase https://doi.org/10.1002/anie.200601815} {\bibfield  {journal}
  {\bibinfo  {journal} {Angew. Chem. Int. Ed.}\ }\textbf {\bibinfo {volume}
  {46}},\ \bibinfo {pages} {668} (\bibinfo {year} {2007})}\BibitemShut
  {NoStop}%
\bibitem [{\citenamefont {Chappert}\ \emph {et~al.}(2007)\citenamefont
  {Chappert}, \citenamefont {Fert},\ and\ \citenamefont {van
  Dau}}]{Chappert2007_NatMater_6_813}%
  \BibitemOpen
  \bibfield  {author} {\bibinfo {author} {\bibfnamefont {C.}~\bibnamefont
  {Chappert}}, \bibinfo {author} {\bibfnamefont {A.}~\bibnamefont {Fert}}, \
  and\ \bibinfo {author} {\bibfnamefont {F.~N.}\ \bibnamefont {van Dau}},\
  }\href@noop {} {\bibfield  {journal} {\bibinfo  {journal} {Nat. Mater.}\
  }\textbf {\bibinfo {volume} {6}},\ \bibinfo {pages} {813} (\bibinfo {year}
  {2007})}\BibitemShut {NoStop}%
\bibitem [{\citenamefont {Bader}\ and\ \citenamefont
  {Parkin}(2010)}]{Bader2010_AnnuRevCondensMatterPhys_1_71}%
  \BibitemOpen
  \bibfield  {author} {\bibinfo {author} {\bibfnamefont {S.}~\bibnamefont
  {Bader}}\ and\ \bibinfo {author} {\bibfnamefont {S.}~\bibnamefont {Parkin}},\
  }\href {\doibase 10.1146/annurev-conmatphys-070909-104123} {\bibfield
  {journal} {\bibinfo  {journal} {Annu. Rev. Condens. Matter Phys.}\ }\textbf
  {\bibinfo {volume} {1}},\ \bibinfo {pages} {71} (\bibinfo {year}
  {2010})}\BibitemShut {NoStop}%
\bibitem [{\citenamefont {Hirohata}\ \emph {et~al.}(2020)\citenamefont
  {Hirohata}, \citenamefont {Yamada}, \citenamefont {Nakatani}, \citenamefont
  {Prejbeanu}, \citenamefont {Di\'{e}ny}, \citenamefont {Pirro},\ and\
  \citenamefont {Hillebrands}}]{Hirohata2020_JMagnMagnMater_509_166711}%
  \BibitemOpen
  \bibfield  {author} {\bibinfo {author} {\bibfnamefont {A.}~\bibnamefont
  {Hirohata}}, \bibinfo {author} {\bibfnamefont {K.}~\bibnamefont {Yamada}},
  \bibinfo {author} {\bibfnamefont {Y.}~\bibnamefont {Nakatani}}, \bibinfo
  {author} {\bibfnamefont {I.-L.}\ \bibnamefont {Prejbeanu}}, \bibinfo {author}
  {\bibfnamefont {B.}~\bibnamefont {Di\'{e}ny}}, \bibinfo {author}
  {\bibfnamefont {P.}~\bibnamefont {Pirro}}, \ and\ \bibinfo {author}
  {\bibfnamefont {B.}~\bibnamefont {Hillebrands}},\ }\href {\doibase
  https://doi.org/10.1016/j.jmmm.2020.166711} {\bibfield  {journal} {\bibinfo
  {journal} {J. Magn. Magn. Mater.}\ }\textbf {\bibinfo {volume} {509}},\
  \bibinfo {pages} {166711} (\bibinfo {year} {2020})}\BibitemShut {NoStop}%
\bibitem [{\citenamefont {Julliere}(1975)}]{Julliere1975_PhysLettA_54A_225}%
  \BibitemOpen
  \bibfield  {author} {\bibinfo {author} {\bibfnamefont {M.}~\bibnamefont
  {Julliere}},\ }\href {\doibase https://doi.org/10.1016/0375-9601(75)90174-7}
  {\bibfield  {journal} {\bibinfo  {journal} {Phys. Lett. A}\ }\textbf
  {\bibinfo {volume} {54}},\ \bibinfo {pages} {225} (\bibinfo {year}
  {1975})}\BibitemShut {NoStop}%
\bibitem [{\citenamefont {{Beth
  Stearns}}(1977)}]{Stearns1977_JMagnMagnMater_5_167}%
  \BibitemOpen
  \bibfield  {author} {\bibinfo {author} {\bibfnamefont {M.}~\bibnamefont
  {{Beth Stearns}}},\ }\href {\doibase
  https://doi.org/10.1016/0304-8853(77)90185-8} {\bibfield  {journal} {\bibinfo
   {journal} {J. Magn. Magn. Mater.}\ }\textbf {\bibinfo {volume} {5}},\
  \bibinfo {pages} {167} (\bibinfo {year} {1977})}\BibitemShut {NoStop}%
\bibitem [{\citenamefont {Tsymbal}\ \emph {et~al.}(2003)\citenamefont
  {Tsymbal}, \citenamefont {Mryasov},\ and\ \citenamefont
  {LeClair}}]{Tsymbal2003_JPhysCondensMatter_15_R109}%
  \BibitemOpen
  \bibfield  {author} {\bibinfo {author} {\bibfnamefont {E.~Y.}\ \bibnamefont
  {Tsymbal}}, \bibinfo {author} {\bibfnamefont {O.~N.}\ \bibnamefont
  {Mryasov}}, \ and\ \bibinfo {author} {\bibfnamefont {P.~R.}\ \bibnamefont
  {LeClair}},\ }\href {\doibase 10.1088/0953-8984/15/4/201} {\bibfield
  {journal} {\bibinfo  {journal} {J. Phys.: Condens. Matter}\ }\textbf
  {\bibinfo {volume} {15}},\ \bibinfo {pages} {R109} (\bibinfo {year}
  {2003})}\BibitemShut {NoStop}%
\bibitem [{\citenamefont {Zhang}\ and\ \citenamefont
  {Butler}(2003)}]{Zhang2003_JPhysCondensMatter_15_R1603}%
  \BibitemOpen
  \bibfield  {author} {\bibinfo {author} {\bibfnamefont {X.-G.}\ \bibnamefont
  {Zhang}}\ and\ \bibinfo {author} {\bibfnamefont {W.~H.}\ \bibnamefont
  {Butler}},\ }\href {\doibase 10.1088/0953-8984/15/41/r01} {\bibfield
  {journal} {\bibinfo  {journal} {J. Phys.: Condens. Matter}\ }\textbf
  {\bibinfo {volume} {15}},\ \bibinfo {pages} {R1603} (\bibinfo {year}
  {2003})}\BibitemShut {NoStop}%
\bibitem [{\citenamefont {Itoh}\ and\ \citenamefont
  {Inoue}(2006)}]{Ito2006_JMagnSocJpn_30_1}%
  \BibitemOpen
  \bibfield  {author} {\bibinfo {author} {\bibfnamefont {H.}~\bibnamefont
  {Itoh}}\ and\ \bibinfo {author} {\bibfnamefont {J.}~\bibnamefont {Inoue}},\
  }\href {\doibase 10.3379/jmsjmag.30.1} {\bibfield  {journal} {\bibinfo
  {journal} {J. Magn. Soc. Jpn.}\ }\textbf {\bibinfo {volume} {30}},\ \bibinfo
  {pages} {1} (\bibinfo {year} {2006})}\BibitemShut {NoStop}%
\bibitem [{\citenamefont {Yuasa}\ and\ \citenamefont
  {Djayaprawira}(2007)}]{Yuasa2007_JPhysD_40_R337}%
  \BibitemOpen
  \bibfield  {author} {\bibinfo {author} {\bibfnamefont {S.}~\bibnamefont
  {Yuasa}}\ and\ \bibinfo {author} {\bibfnamefont {D.~D.}\ \bibnamefont
  {Djayaprawira}},\ }\href {\doibase 10.1088/0022-3727/40/21/r01} {\bibfield
  {journal} {\bibinfo  {journal} {J. Phys. D: Appl. Phys.}\ }\textbf {\bibinfo
  {volume} {40}},\ \bibinfo {pages} {R337} (\bibinfo {year}
  {2007})}\BibitemShut {NoStop}%
\bibitem [{\citenamefont
  {Butler}(2008)}]{Butler2008_SciTechnolAdvMater_9_014106}%
  \BibitemOpen
  \bibfield  {author} {\bibinfo {author} {\bibfnamefont {W.~H.}\ \bibnamefont
  {Butler}},\ }\href {\doibase 10.1088/1468-6996/9/1/014106} {\bibfield
  {journal} {\bibinfo  {journal} {Sci. Technol. Adv. Mater.}\ }\textbf
  {\bibinfo {volume} {9}},\ \bibinfo {pages} {014106} (\bibinfo {year}
  {2008})}\BibitemShut {NoStop}%
\bibitem [{\citenamefont
  {Slonczewski}(1989)}]{Slonczewski1989_PhysRevB_39_6995}%
  \BibitemOpen
  \bibfield  {author} {\bibinfo {author} {\bibfnamefont {J.~C.}\ \bibnamefont
  {Slonczewski}},\ }\href {\doibase 10.1103/PhysRevB.39.6995} {\bibfield
  {journal} {\bibinfo  {journal} {Phys. Rev. B}\ }\textbf {\bibinfo {volume}
  {39}},\ \bibinfo {pages} {6995} (\bibinfo {year} {1989})}\BibitemShut
  {NoStop}%
\bibitem [{\citenamefont {Mathon}(1997)}]{Mathon1997_PhysRevB_56_11810}%
  \BibitemOpen
  \bibfield  {author} {\bibinfo {author} {\bibfnamefont {J.}~\bibnamefont
  {Mathon}},\ }\href {\doibase 10.1103/PhysRevB.56.11810} {\bibfield  {journal}
  {\bibinfo  {journal} {Phys. Rev. B}\ }\textbf {\bibinfo {volume} {56}},\
  \bibinfo {pages} {11810} (\bibinfo {year} {1997})}\BibitemShut {NoStop}%
\bibitem [{\citenamefont {Butler}\ \emph {et~al.}(2001)\citenamefont {Butler},
  \citenamefont {Zhang}, \citenamefont {Schulthess},\ and\ \citenamefont
  {MacLaren}}]{Butler2001_PhysRevB_63_054416}%
  \BibitemOpen
  \bibfield  {author} {\bibinfo {author} {\bibfnamefont {W.~H.}\ \bibnamefont
  {Butler}}, \bibinfo {author} {\bibfnamefont {X.-G.}\ \bibnamefont {Zhang}},
  \bibinfo {author} {\bibfnamefont {T.~C.}\ \bibnamefont {Schulthess}}, \ and\
  \bibinfo {author} {\bibfnamefont {J.~M.}\ \bibnamefont {MacLaren}},\ }\href
  {\doibase 10.1103/PhysRevB.63.054416} {\bibfield  {journal} {\bibinfo
  {journal} {Phys. Rev. B}\ }\textbf {\bibinfo {volume} {63}},\ \bibinfo
  {pages} {054416} (\bibinfo {year} {2001})}\BibitemShut {NoStop}%
\bibitem [{\citenamefont {Mathon}\ and\ \citenamefont
  {Umerski}(2001)}]{Mathon2001_PhysRevB_63_220403}%
  \BibitemOpen
  \bibfield  {author} {\bibinfo {author} {\bibfnamefont {J.}~\bibnamefont
  {Mathon}}\ and\ \bibinfo {author} {\bibfnamefont {A.}~\bibnamefont
  {Umerski}},\ }\href {\doibase 10.1103/PhysRevB.63.220403} {\bibfield
  {journal} {\bibinfo  {journal} {Phys. Rev. B}\ }\textbf {\bibinfo {volume}
  {63}},\ \bibinfo {pages} {220403} (\bibinfo {year} {2001})}\BibitemShut
  {NoStop}%
\bibitem [{\citenamefont {Miyazaki}\ and\ \citenamefont
  {Tezuka}(1995)}]{Miyazaki1995_JMagnMagnMater_139_L231}%
  \BibitemOpen
  \bibfield  {author} {\bibinfo {author} {\bibfnamefont {T.}~\bibnamefont
  {Miyazaki}}\ and\ \bibinfo {author} {\bibfnamefont {N.}~\bibnamefont
  {Tezuka}},\ }\href {\doibase https://doi.org/10.1016/0304-8853(95)90001-2}
  {\bibfield  {journal} {\bibinfo  {journal} {J. Magn. Magn. Mater.}\ }\textbf
  {\bibinfo {volume} {139}},\ \bibinfo {pages} {L231} (\bibinfo {year}
  {1995})}\BibitemShut {NoStop}%
\bibitem [{\citenamefont {Moodera}\ \emph {et~al.}(1995)\citenamefont
  {Moodera}, \citenamefont {Kinder}, \citenamefont {Wong},\ and\ \citenamefont
  {Meservey}}]{Moodera1995_PhysRevLett_74_3273}%
  \BibitemOpen
  \bibfield  {author} {\bibinfo {author} {\bibfnamefont {J.~S.}\ \bibnamefont
  {Moodera}}, \bibinfo {author} {\bibfnamefont {L.~R.}\ \bibnamefont {Kinder}},
  \bibinfo {author} {\bibfnamefont {T.~M.}\ \bibnamefont {Wong}}, \ and\
  \bibinfo {author} {\bibfnamefont {R.}~\bibnamefont {Meservey}},\ }\href
  {\doibase 10.1103/PhysRevLett.74.3273} {\bibfield  {journal} {\bibinfo
  {journal} {Phys. Rev. Lett.}\ }\textbf {\bibinfo {volume} {74}},\ \bibinfo
  {pages} {3273} (\bibinfo {year} {1995})}\BibitemShut {NoStop}%
\bibitem [{\citenamefont {Yuasa}\ \emph {et~al.}(2004)\citenamefont {Yuasa},
  \citenamefont {Nagahama}, \citenamefont {Fukushima}, \citenamefont {Suzuki},\
  and\ \citenamefont {Ando}}]{Yuasa2004_NatMater_3_868}%
  \BibitemOpen
  \bibfield  {author} {\bibinfo {author} {\bibfnamefont {S.}~\bibnamefont
  {Yuasa}}, \bibinfo {author} {\bibfnamefont {T.}~\bibnamefont {Nagahama}},
  \bibinfo {author} {\bibfnamefont {A.}~\bibnamefont {Fukushima}}, \bibinfo
  {author} {\bibfnamefont {Y.}~\bibnamefont {Suzuki}}, \ and\ \bibinfo {author}
  {\bibfnamefont {K.}~\bibnamefont {Ando}},\ }\href@noop {} {\bibfield
  {journal} {\bibinfo  {journal} {Nat. Mater.}\ }\textbf {\bibinfo {volume}
  {3}},\ \bibinfo {pages} {868} (\bibinfo {year} {2004})}\BibitemShut {NoStop}%
\bibitem [{\citenamefont {Parkin}\ \emph {et~al.}(2004)\citenamefont {Parkin},
  \citenamefont {Kaiser}, \citenamefont {Panchula}, \citenamefont {Rice},
  \citenamefont {Hughes}, \citenamefont {Samant},\ and\ \citenamefont
  {Yang}}]{Parkin2004_NatMater_3_862}%
  \BibitemOpen
  \bibfield  {author} {\bibinfo {author} {\bibfnamefont {S.~S.}\ \bibnamefont
  {Parkin}}, \bibinfo {author} {\bibfnamefont {C.}~\bibnamefont {Kaiser}},
  \bibinfo {author} {\bibfnamefont {A.}~\bibnamefont {Panchula}}, \bibinfo
  {author} {\bibfnamefont {P.~M.}\ \bibnamefont {Rice}}, \bibinfo {author}
  {\bibfnamefont {B.}~\bibnamefont {Hughes}}, \bibinfo {author} {\bibfnamefont
  {M.}~\bibnamefont {Samant}}, \ and\ \bibinfo {author} {\bibfnamefont {S.-H.}\
  \bibnamefont {Yang}},\ }\href@noop {} {\bibfield  {journal} {\bibinfo
  {journal} {Nat. Mater.}\ }\textbf {\bibinfo {volume} {3}},\ \bibinfo {pages}
  {862} (\bibinfo {year} {2004})}\BibitemShut {NoStop}%
\bibitem [{\citenamefont {Djayaprawira}\ \emph {et~al.}(2005)\citenamefont
  {Djayaprawira}, \citenamefont {Tsunekawa}, \citenamefont {Nagai},
  \citenamefont {Maehara}, \citenamefont {Yamagata}, \citenamefont {Watanabe},
  \citenamefont {Yuasa}, \citenamefont {Suzuki},\ and\ \citenamefont
  {Ando}}]{Djayaprawira2005_ApplPhysLett_86_092502}%
  \BibitemOpen
  \bibfield  {author} {\bibinfo {author} {\bibfnamefont {D.~D.}\ \bibnamefont
  {Djayaprawira}}, \bibinfo {author} {\bibfnamefont {K.}~\bibnamefont
  {Tsunekawa}}, \bibinfo {author} {\bibfnamefont {M.}~\bibnamefont {Nagai}},
  \bibinfo {author} {\bibfnamefont {H.}~\bibnamefont {Maehara}}, \bibinfo
  {author} {\bibfnamefont {S.}~\bibnamefont {Yamagata}}, \bibinfo {author}
  {\bibfnamefont {N.}~\bibnamefont {Watanabe}}, \bibinfo {author}
  {\bibfnamefont {S.}~\bibnamefont {Yuasa}}, \bibinfo {author} {\bibfnamefont
  {Y.}~\bibnamefont {Suzuki}}, \ and\ \bibinfo {author} {\bibfnamefont
  {K.}~\bibnamefont {Ando}},\ }\href {\doibase 10.1063/1.1871344} {\bibfield
  {journal} {\bibinfo  {journal} {Appl. Phys. Lett.}\ }\textbf {\bibinfo
  {volume} {86}},\ \bibinfo {pages} {092502} (\bibinfo {year}
  {2005})}\BibitemShut {NoStop}%
\bibitem [{\citenamefont {Ikeda}\ \emph {et~al.}(2008)\citenamefont {Ikeda},
  \citenamefont {Hayakawa}, \citenamefont {Ashizawa}, \citenamefont {Lee},
  \citenamefont {Miura}, \citenamefont {Hasegawa}, \citenamefont {Tsunoda},
  \citenamefont {Matsukura},\ and\ \citenamefont
  {Ohno}}]{Ikeda2008_ApplPhysLett_93_082508}%
  \BibitemOpen
  \bibfield  {author} {\bibinfo {author} {\bibfnamefont {S.}~\bibnamefont
  {Ikeda}}, \bibinfo {author} {\bibfnamefont {J.}~\bibnamefont {Hayakawa}},
  \bibinfo {author} {\bibfnamefont {Y.}~\bibnamefont {Ashizawa}}, \bibinfo
  {author} {\bibfnamefont {Y.~M.}\ \bibnamefont {Lee}}, \bibinfo {author}
  {\bibfnamefont {K.}~\bibnamefont {Miura}}, \bibinfo {author} {\bibfnamefont
  {H.}~\bibnamefont {Hasegawa}}, \bibinfo {author} {\bibfnamefont
  {M.}~\bibnamefont {Tsunoda}}, \bibinfo {author} {\bibfnamefont
  {F.}~\bibnamefont {Matsukura}}, \ and\ \bibinfo {author} {\bibfnamefont
  {H.}~\bibnamefont {Ohno}},\ }\href {\doibase 10.1063/1.2976435} {\bibfield
  {journal} {\bibinfo  {journal} {Appl. Phys. Lett.}\ }\textbf {\bibinfo
  {volume} {93}},\ \bibinfo {pages} {082508} (\bibinfo {year}
  {2008})}\BibitemShut {NoStop}%
\bibitem [{\citenamefont {Tanaka}\ \emph {et~al.}(1999)\citenamefont {Tanaka},
  \citenamefont {Nowak},\ and\ \citenamefont
  {Moodera}}]{Tanaka1999_JApplPhys_86_6239}%
  \BibitemOpen
  \bibfield  {author} {\bibinfo {author} {\bibfnamefont {C.~T.}\ \bibnamefont
  {Tanaka}}, \bibinfo {author} {\bibfnamefont {J.}~\bibnamefont {Nowak}}, \
  and\ \bibinfo {author} {\bibfnamefont {J.~S.}\ \bibnamefont {Moodera}},\
  }\href {\doibase 10.1063/1.371678} {\bibfield  {journal} {\bibinfo  {journal}
  {J. Appl. Phys.}\ }\textbf {\bibinfo {volume} {86}},\ \bibinfo {pages} {6239}
  (\bibinfo {year} {1999})}\BibitemShut {NoStop}%
\bibitem [{\citenamefont {Inomata}\ \emph {et~al.}(2003)\citenamefont
  {Inomata}, \citenamefont {Okamura}, \citenamefont {Goto},\ and\ \citenamefont
  {Tezuka}}]{Inomata2003_JpnJApplPhys_42_L419}%
  \BibitemOpen
  \bibfield  {author} {\bibinfo {author} {\bibfnamefont {K.}~\bibnamefont
  {Inomata}}, \bibinfo {author} {\bibfnamefont {S.}~\bibnamefont {Okamura}},
  \bibinfo {author} {\bibfnamefont {R.}~\bibnamefont {Goto}}, \ and\ \bibinfo
  {author} {\bibfnamefont {N.}~\bibnamefont {Tezuka}},\ }\href {\doibase
  10.1143/jjap.42.l419} {\bibfield  {journal} {\bibinfo  {journal} {Jpn. J.
  Appl. Phys.}\ }\textbf {\bibinfo {volume} {42}},\ \bibinfo {pages} {L419}
  (\bibinfo {year} {2003})}\BibitemShut {NoStop}%
\bibitem [{\citenamefont {K\"{a}mmerer}\ \emph {et~al.}(2004)\citenamefont
  {K\"{a}mmerer}, \citenamefont {Thomas}, \citenamefont {H\"{u}tten},\ and\
  \citenamefont {Reiss}}]{Kammerer2004_ApplPhysLett_85_79}%
  \BibitemOpen
  \bibfield  {author} {\bibinfo {author} {\bibfnamefont {S.}~\bibnamefont
  {K\"{a}mmerer}}, \bibinfo {author} {\bibfnamefont {A.}~\bibnamefont
  {Thomas}}, \bibinfo {author} {\bibfnamefont {A.}~\bibnamefont {H\"{u}tten}},
  \ and\ \bibinfo {author} {\bibfnamefont {G.}~\bibnamefont {Reiss}},\ }\href
  {\doibase 10.1063/1.1769082} {\bibfield  {journal} {\bibinfo  {journal}
  {Appl. Phys. Lett.}\ }\textbf {\bibinfo {volume} {85}},\ \bibinfo {pages}
  {79} (\bibinfo {year} {2004})}\BibitemShut {NoStop}%
\bibitem [{\citenamefont {Liu}\ \emph {et~al.}(2012)\citenamefont {Liu},
  \citenamefont {Honda}, \citenamefont {Taira}, \citenamefont {Matsuda},
  \citenamefont {Arita}, \citenamefont {Uemura},\ and\ \citenamefont
  {Yamamoto}}]{Liu2012_ApplPhysLett_101_132418}%
  \BibitemOpen
  \bibfield  {author} {\bibinfo {author} {\bibfnamefont {H.-x.}\ \bibnamefont
  {Liu}}, \bibinfo {author} {\bibfnamefont {Y.}~\bibnamefont {Honda}}, \bibinfo
  {author} {\bibfnamefont {T.}~\bibnamefont {Taira}}, \bibinfo {author}
  {\bibfnamefont {K.-i.}\ \bibnamefont {Matsuda}}, \bibinfo {author}
  {\bibfnamefont {M.}~\bibnamefont {Arita}}, \bibinfo {author} {\bibfnamefont
  {T.}~\bibnamefont {Uemura}}, \ and\ \bibinfo {author} {\bibfnamefont
  {M.}~\bibnamefont {Yamamoto}},\ }\href {\doibase 10.1063/1.4755773}
  {\bibfield  {journal} {\bibinfo  {journal} {Appl. Phys. Lett.}\ }\textbf
  {\bibinfo {volume} {101}},\ \bibinfo {pages} {132418} (\bibinfo {year}
  {2012})}\BibitemShut {NoStop}%
\bibitem [{\citenamefont {MacDonald}\ and\ \citenamefont
  {Tsoi}(2011)}]{MacDonald2011_PhilTransRSocA_369_3098}%
  \BibitemOpen
  \bibfield  {author} {\bibinfo {author} {\bibfnamefont {A.~H.}\ \bibnamefont
  {MacDonald}}\ and\ \bibinfo {author} {\bibfnamefont {M.}~\bibnamefont
  {Tsoi}},\ }\href {\doibase 10.1098/rsta.2011.0014} {\bibfield  {journal}
  {\bibinfo  {journal} {Phil. Trans. R. Soc. A}\ }\textbf {\bibinfo {volume}
  {369}},\ \bibinfo {pages} {3098} (\bibinfo {year} {2011})}\BibitemShut
  {NoStop}%
\bibitem [{\citenamefont {Gomonay}\ and\ \citenamefont
  {Loktev}(2014)}]{Gomonay2014_LowTempPhys_40_17}%
  \BibitemOpen
  \bibfield  {author} {\bibinfo {author} {\bibfnamefont {E.~V.}\ \bibnamefont
  {Gomonay}}\ and\ \bibinfo {author} {\bibfnamefont {V.~M.}\ \bibnamefont
  {Loktev}},\ }\href {\doibase 10.1063/1.4862467} {\bibfield  {journal}
  {\bibinfo  {journal} {Low Temp. Phys.}\ }\textbf {\bibinfo {volume} {40}},\
  \bibinfo {pages} {17} (\bibinfo {year} {2014})}\BibitemShut {NoStop}%
\bibitem [{\citenamefont {Jungwirth}\ \emph {et~al.}(2016)\citenamefont
  {Jungwirth}, \citenamefont {Marti}, \citenamefont {Wadley},\ and\
  \citenamefont {Wunderlich}}]{Jungwirth2016_NatNanotechnol_11_231}%
  \BibitemOpen
  \bibfield  {author} {\bibinfo {author} {\bibfnamefont {T.}~\bibnamefont
  {Jungwirth}}, \bibinfo {author} {\bibfnamefont {X.}~\bibnamefont {Marti}},
  \bibinfo {author} {\bibfnamefont {P.}~\bibnamefont {Wadley}}, \ and\ \bibinfo
  {author} {\bibfnamefont {J.}~\bibnamefont {Wunderlich}},\ }\href@noop {}
  {\bibfield  {journal} {\bibinfo  {journal} {Nat. Nanotechnol.}\ }\textbf
  {\bibinfo {volume} {11}},\ \bibinfo {pages} {231} (\bibinfo {year}
  {2016})}\BibitemShut {NoStop}%
\bibitem [{\citenamefont {Baltz}\ \emph {et~al.}(2018)\citenamefont {Baltz},
  \citenamefont {Manchon}, \citenamefont {Tsoi}, \citenamefont {Moriyama},
  \citenamefont {Ono},\ and\ \citenamefont
  {Tserkovnyak}}]{Baltz2018_RevModPhys_90_015005}%
  \BibitemOpen
  \bibfield  {author} {\bibinfo {author} {\bibfnamefont {V.}~\bibnamefont
  {Baltz}}, \bibinfo {author} {\bibfnamefont {A.}~\bibnamefont {Manchon}},
  \bibinfo {author} {\bibfnamefont {M.}~\bibnamefont {Tsoi}}, \bibinfo {author}
  {\bibfnamefont {T.}~\bibnamefont {Moriyama}}, \bibinfo {author}
  {\bibfnamefont {T.}~\bibnamefont {Ono}}, \ and\ \bibinfo {author}
  {\bibfnamefont {Y.}~\bibnamefont {Tserkovnyak}},\ }\href {\doibase
  10.1103/RevModPhys.90.015005} {\bibfield  {journal} {\bibinfo  {journal}
  {Rev. Mod. Phys.}\ }\textbf {\bibinfo {volume} {90}},\ \bibinfo {pages}
  {015005} (\bibinfo {year} {2018})}\BibitemShut {NoStop}%
\bibitem [{\citenamefont {{\v{Z}}elezn{\`y}}\ \emph {et~al.}(2018)\citenamefont
  {{\v{Z}}elezn{\`y}}, \citenamefont {Wadley}, \citenamefont {Olejn{\'\i}k},
  \citenamefont {Hoffmann},\ and\ \citenamefont
  {Ohno}}]{Zelezny2018_NatPhys_14_220}%
  \BibitemOpen
  \bibfield  {author} {\bibinfo {author} {\bibfnamefont {J.}~\bibnamefont
  {{\v{Z}}elezn{\`y}}}, \bibinfo {author} {\bibfnamefont {P.}~\bibnamefont
  {Wadley}}, \bibinfo {author} {\bibfnamefont {K.}~\bibnamefont
  {Olejn{\'\i}k}}, \bibinfo {author} {\bibfnamefont {A.}~\bibnamefont
  {Hoffmann}}, \ and\ \bibinfo {author} {\bibfnamefont {H.}~\bibnamefont
  {Ohno}},\ }\href@noop {} {\bibfield  {journal} {\bibinfo  {journal} {Nat.
  Phys.}\ }\textbf {\bibinfo {volume} {14}},\ \bibinfo {pages} {220} (\bibinfo
  {year} {2018})}\BibitemShut {NoStop}%
\bibitem [{\citenamefont {Siddiqui}\ \emph {et~al.}(2020)\citenamefont
  {Siddiqui}, \citenamefont {Sklenar}, \citenamefont {Kang}, \citenamefont
  {Gilbert}, \citenamefont {Schleife}, \citenamefont {Mason},\ and\
  \citenamefont {Hoffmann}}]{Siddiqui2020_JApplPhys_128_040904}%
  \BibitemOpen
  \bibfield  {author} {\bibinfo {author} {\bibfnamefont {S.~A.}\ \bibnamefont
  {Siddiqui}}, \bibinfo {author} {\bibfnamefont {J.}~\bibnamefont {Sklenar}},
  \bibinfo {author} {\bibfnamefont {K.}~\bibnamefont {Kang}}, \bibinfo {author}
  {\bibfnamefont {M.~J.}\ \bibnamefont {Gilbert}}, \bibinfo {author}
  {\bibfnamefont {A.}~\bibnamefont {Schleife}}, \bibinfo {author}
  {\bibfnamefont {N.}~\bibnamefont {Mason}}, \ and\ \bibinfo {author}
  {\bibfnamefont {A.}~\bibnamefont {Hoffmann}},\ }\href {\doibase
  10.1063/5.0009445} {\bibfield  {journal} {\bibinfo  {journal} {J. Appl.
  Phys.}\ }\textbf {\bibinfo {volume} {128}},\ \bibinfo {pages} {040904}
  (\bibinfo {year} {2020})}\BibitemShut {NoStop}%
\bibitem [{\citenamefont {Fukami}\ \emph {et~al.}(2020)\citenamefont {Fukami},
  \citenamefont {Lorenz},\ and\ \citenamefont
  {Gomonay}}]{Fukami2020_JApplPhys_128_070401}%
  \BibitemOpen
  \bibfield  {author} {\bibinfo {author} {\bibfnamefont {S.}~\bibnamefont
  {Fukami}}, \bibinfo {author} {\bibfnamefont {V.~O.}\ \bibnamefont {Lorenz}},
  \ and\ \bibinfo {author} {\bibfnamefont {O.}~\bibnamefont {Gomonay}},\ }\href
  {\doibase 10.1063/5.0023614} {\bibfield  {journal} {\bibinfo  {journal} {J.
  Appl. Phys.}\ }\textbf {\bibinfo {volume} {128}},\ \bibinfo {pages} {070401}
  (\bibinfo {year} {2020})}\BibitemShut {NoStop}%
\bibitem [{\citenamefont {Barker}\ and\ \citenamefont
  {Atxitia}(2021)}]{Barker2021_JPhysSocJpn_90_081001}%
  \BibitemOpen
  \bibfield  {author} {\bibinfo {author} {\bibfnamefont {J.}~\bibnamefont
  {Barker}}\ and\ \bibinfo {author} {\bibfnamefont {U.}~\bibnamefont
  {Atxitia}},\ }\href {\doibase 10.7566/JPSJ.90.081001} {\bibfield  {journal}
  {\bibinfo  {journal} {J. Phys. Soc. Jpn.}\ }\textbf {\bibinfo {volume}
  {90}},\ \bibinfo {pages} {081001} (\bibinfo {year} {2021})}\BibitemShut
  {NoStop}%
\bibitem [{\citenamefont {Kim}\ \emph {et~al.}(2022)\citenamefont {Kim},
  \citenamefont {Beach}, \citenamefont {Lee}, \citenamefont {Ono},
  \citenamefont {Rasing},\ and\ \citenamefont {Yang}}]{Kim2022_NatMater_21_24}%
  \BibitemOpen
  \bibfield  {author} {\bibinfo {author} {\bibfnamefont {S.~K.}\ \bibnamefont
  {Kim}}, \bibinfo {author} {\bibfnamefont {G.~S.}\ \bibnamefont {Beach}},
  \bibinfo {author} {\bibfnamefont {K.-J.}\ \bibnamefont {Lee}}, \bibinfo
  {author} {\bibfnamefont {T.}~\bibnamefont {Ono}}, \bibinfo {author}
  {\bibfnamefont {T.}~\bibnamefont {Rasing}}, \ and\ \bibinfo {author}
  {\bibfnamefont {H.}~\bibnamefont {Yang}},\ }\href@noop {} {\bibfield
  {journal} {\bibinfo  {journal} {Nat. Mater.}\ }\textbf {\bibinfo {volume}
  {21}},\ \bibinfo {pages} {24} (\bibinfo {year} {2022})}\BibitemShut {NoStop}%
\bibitem [{\citenamefont {N\'u\~nez}\ \emph {et~al.}(2006)\citenamefont
  {N\'u\~nez}, \citenamefont {Duine}, \citenamefont {Haney},\ and\
  \citenamefont {MacDonald}}]{Nunez2006_PhysRevB_73_214426}%
  \BibitemOpen
  \bibfield  {author} {\bibinfo {author} {\bibfnamefont {A.~S.}\ \bibnamefont
  {N\'u\~nez}}, \bibinfo {author} {\bibfnamefont {R.~A.}\ \bibnamefont
  {Duine}}, \bibinfo {author} {\bibfnamefont {P.}~\bibnamefont {Haney}}, \ and\
  \bibinfo {author} {\bibfnamefont {A.~H.}\ \bibnamefont {MacDonald}},\ }\href
  {\doibase 10.1103/PhysRevB.73.214426} {\bibfield  {journal} {\bibinfo
  {journal} {Phys. Rev. B}\ }\textbf {\bibinfo {volume} {73}},\ \bibinfo
  {pages} {214426} (\bibinfo {year} {2006})}\BibitemShut {NoStop}%
\bibitem [{\citenamefont {Saidaoui}\ \emph {et~al.}(2014)\citenamefont
  {Saidaoui}, \citenamefont {Manchon},\ and\ \citenamefont
  {Waintal}}]{Saidaoui2014_PhysRevB_89_174430}%
  \BibitemOpen
  \bibfield  {author} {\bibinfo {author} {\bibfnamefont {H.~B.~M.}\
  \bibnamefont {Saidaoui}}, \bibinfo {author} {\bibfnamefont {A.}~\bibnamefont
  {Manchon}}, \ and\ \bibinfo {author} {\bibfnamefont {X.}~\bibnamefont
  {Waintal}},\ }\href {\doibase 10.1103/PhysRevB.89.174430} {\bibfield
  {journal} {\bibinfo  {journal} {Phys. Rev. B}\ }\textbf {\bibinfo {volume}
  {89}},\ \bibinfo {pages} {174430} (\bibinfo {year} {2014})}\BibitemShut
  {NoStop}%
\bibitem [{\citenamefont {Ghosh}\ \emph {et~al.}(2022)\citenamefont {Ghosh},
  \citenamefont {Manchon},\ and\ \citenamefont {\ifmmode~\check{Z}\else
  \v{Z}\fi{}elezn\'y}}]{Ghosh2022_PhysRevLett_128_097702}%
  \BibitemOpen
  \bibfield  {author} {\bibinfo {author} {\bibfnamefont {S.}~\bibnamefont
  {Ghosh}}, \bibinfo {author} {\bibfnamefont {A.}~\bibnamefont {Manchon}}, \
  and\ \bibinfo {author} {\bibfnamefont {J.}~\bibnamefont
  {\ifmmode~\check{Z}\else \v{Z}\fi{}elezn\'y}},\ }\href {\doibase
  10.1103/PhysRevLett.128.097702} {\bibfield  {journal} {\bibinfo  {journal}
  {Phys. Rev. Lett.}\ }\textbf {\bibinfo {volume} {128}},\ \bibinfo {pages}
  {097702} (\bibinfo {year} {2022})}\BibitemShut {NoStop}%
\bibitem [{\citenamefont {Chen}\ \emph {et~al.}(2014)\citenamefont {Chen},
  \citenamefont {Niu},\ and\ \citenamefont
  {MacDonald}}]{Chen2014_PhysRevLett_112_017205}%
  \BibitemOpen
  \bibfield  {author} {\bibinfo {author} {\bibfnamefont {H.}~\bibnamefont
  {Chen}}, \bibinfo {author} {\bibfnamefont {Q.}~\bibnamefont {Niu}}, \ and\
  \bibinfo {author} {\bibfnamefont {A.~H.}\ \bibnamefont {MacDonald}},\ }\href
  {\doibase 10.1103/PhysRevLett.112.017205} {\bibfield  {journal} {\bibinfo
  {journal} {Phys. Rev. Lett.}\ }\textbf {\bibinfo {volume} {112}},\ \bibinfo
  {pages} {017205} (\bibinfo {year} {2014})}\BibitemShut {NoStop}%
\bibitem [{\citenamefont {Nakatsuji}\ \emph {et~al.}(2015)\citenamefont
  {Nakatsuji}, \citenamefont {Kiyohara},\ and\ \citenamefont
  {Higo}}]{Nakatsuji2015_Nature_527_212}%
  \BibitemOpen
  \bibfield  {author} {\bibinfo {author} {\bibfnamefont {S.}~\bibnamefont
  {Nakatsuji}}, \bibinfo {author} {\bibfnamefont {N.}~\bibnamefont {Kiyohara}},
  \ and\ \bibinfo {author} {\bibfnamefont {T.}~\bibnamefont {Higo}},\
  }\href@noop {} {\bibfield  {journal} {\bibinfo  {journal} {Nature}\ }\textbf
  {\bibinfo {volume} {527}},\ \bibinfo {pages} {212} (\bibinfo {year}
  {2015})}\BibitemShut {NoStop}%
\bibitem [{\citenamefont {Kiyohara}\ \emph {et~al.}(2016)\citenamefont
  {Kiyohara}, \citenamefont {Tomita},\ and\ \citenamefont
  {Nakatsuji}}]{Kiyohara2016_PhysRevApplied_5_064009}%
  \BibitemOpen
  \bibfield  {author} {\bibinfo {author} {\bibfnamefont {N.}~\bibnamefont
  {Kiyohara}}, \bibinfo {author} {\bibfnamefont {T.}~\bibnamefont {Tomita}}, \
  and\ \bibinfo {author} {\bibfnamefont {S.}~\bibnamefont {Nakatsuji}},\ }\href
  {\doibase 10.1103/PhysRevApplied.5.064009} {\bibfield  {journal} {\bibinfo
  {journal} {Phys. Rev. Applied}\ }\textbf {\bibinfo {volume} {5}},\ \bibinfo
  {pages} {064009} (\bibinfo {year} {2016})}\BibitemShut {NoStop}%
\bibitem [{\citenamefont {Nayak}\ \emph {et~al.}(2016)\citenamefont {Nayak},
  \citenamefont {Fischer}, \citenamefont {Sun}, \citenamefont {Yan},
  \citenamefont {Karel}, \citenamefont {Komarek}, \citenamefont {Shekhar},
  \citenamefont {Kumar}, \citenamefont {Schnelle}, \citenamefont {K\"{u}bler},
  \citenamefont {Felser},\ and\ \citenamefont
  {Parkin}}]{Nayak2016_SciAdv_2_e1501870}%
  \BibitemOpen
  \bibfield  {author} {\bibinfo {author} {\bibfnamefont {A.~K.}\ \bibnamefont
  {Nayak}}, \bibinfo {author} {\bibfnamefont {J.~E.}\ \bibnamefont {Fischer}},
  \bibinfo {author} {\bibfnamefont {Y.}~\bibnamefont {Sun}}, \bibinfo {author}
  {\bibfnamefont {B.}~\bibnamefont {Yan}}, \bibinfo {author} {\bibfnamefont
  {J.}~\bibnamefont {Karel}}, \bibinfo {author} {\bibfnamefont {A.~C.}\
  \bibnamefont {Komarek}}, \bibinfo {author} {\bibfnamefont {C.}~\bibnamefont
  {Shekhar}}, \bibinfo {author} {\bibfnamefont {N.}~\bibnamefont {Kumar}},
  \bibinfo {author} {\bibfnamefont {W.}~\bibnamefont {Schnelle}}, \bibinfo
  {author} {\bibfnamefont {J.}~\bibnamefont {K\"{u}bler}}, \bibinfo {author}
  {\bibfnamefont {C.}~\bibnamefont {Felser}}, \ and\ \bibinfo {author}
  {\bibfnamefont {S.~S.~P.}\ \bibnamefont {Parkin}},\ }\href {\doibase
  10.1126/sciadv.1501870} {\bibfield  {journal} {\bibinfo  {journal} {Sci.
  Adv.}\ }\textbf {\bibinfo {volume} {2}},\ \bibinfo {pages} {e1501870}
  (\bibinfo {year} {2016})}\BibitemShut {NoStop}%
\bibitem [{\citenamefont {Merodio}\ \emph {et~al.}(2014)\citenamefont
  {Merodio}, \citenamefont {Kalitsov}, \citenamefont {Béa}, \citenamefont
  {Baltz},\ and\ \citenamefont
  {Chshiev}}]{Merodio2014_ApplPhysLett_105_122403}%
  \BibitemOpen
  \bibfield  {author} {\bibinfo {author} {\bibfnamefont {P.}~\bibnamefont
  {Merodio}}, \bibinfo {author} {\bibfnamefont {A.}~\bibnamefont {Kalitsov}},
  \bibinfo {author} {\bibfnamefont {H.}~\bibnamefont {Béa}}, \bibinfo {author}
  {\bibfnamefont {V.}~\bibnamefont {Baltz}}, \ and\ \bibinfo {author}
  {\bibfnamefont {M.}~\bibnamefont {Chshiev}},\ }\href {\doibase
  10.1063/1.4896291} {\bibfield  {journal} {\bibinfo  {journal} {Appl. Phys.
  Lett.}\ }\textbf {\bibinfo {volume} {105}},\ \bibinfo {pages} {122403}
  (\bibinfo {year} {2014})}\BibitemShut {NoStop}%
\bibitem [{\citenamefont {Jeong}\ \emph {et~al.}(2016)\citenamefont {Jeong},
  \citenamefont {Ferrante}, \citenamefont {Faleev}, \citenamefont {Samant},
  \citenamefont {Felser},\ and\ \citenamefont
  {Parkin}}]{Jeong2016_NatCommun_7_10276}%
  \BibitemOpen
  \bibfield  {author} {\bibinfo {author} {\bibfnamefont {J.}~\bibnamefont
  {Jeong}}, \bibinfo {author} {\bibfnamefont {Y.}~\bibnamefont {Ferrante}},
  \bibinfo {author} {\bibfnamefont {S.~V.}\ \bibnamefont {Faleev}}, \bibinfo
  {author} {\bibfnamefont {M.~G.}\ \bibnamefont {Samant}}, \bibinfo {author}
  {\bibfnamefont {C.}~\bibnamefont {Felser}}, \ and\ \bibinfo {author}
  {\bibfnamefont {S.~S.}\ \bibnamefont {Parkin}},\ }\href@noop {} {\bibfield
  {journal} {\bibinfo  {journal} {Nat. Commun.}\ }\textbf {\bibinfo {volume}
  {7}},\ \bibinfo {pages} {10276} (\bibinfo {year} {2016})}\BibitemShut
  {NoStop}%
\bibitem [{\citenamefont {Saidaoui}\ \emph {et~al.}(2017)\citenamefont
  {Saidaoui}, \citenamefont {Waintal},\ and\ \citenamefont
  {Manchon}}]{Saidaoui2017_PhysRevB_95_134424}%
  \BibitemOpen
  \bibfield  {author} {\bibinfo {author} {\bibfnamefont {H.~B.~M.}\
  \bibnamefont {Saidaoui}}, \bibinfo {author} {\bibfnamefont {X.}~\bibnamefont
  {Waintal}}, \ and\ \bibinfo {author} {\bibfnamefont {A.}~\bibnamefont
  {Manchon}},\ }\href {\doibase 10.1103/PhysRevB.95.134424} {\bibfield
  {journal} {\bibinfo  {journal} {Phys. Rev. B}\ }\textbf {\bibinfo {volume}
  {95}},\ \bibinfo {pages} {134424} (\bibinfo {year} {2017})}\BibitemShut
  {NoStop}%
\bibitem [{\citenamefont {Shao}\ \emph {et~al.}(2021)\citenamefont {Shao},
  \citenamefont {Zhang}, \citenamefont {Li}, \citenamefont {Eom},\ and\
  \citenamefont {Tsymbal}}]{Shao2021_NatCommun_12_7061}%
  \BibitemOpen
  \bibfield  {author} {\bibinfo {author} {\bibfnamefont {D.-F.}\ \bibnamefont
  {Shao}}, \bibinfo {author} {\bibfnamefont {S.-H.}\ \bibnamefont {Zhang}},
  \bibinfo {author} {\bibfnamefont {M.}~\bibnamefont {Li}}, \bibinfo {author}
  {\bibfnamefont {C.-B.}\ \bibnamefont {Eom}}, \ and\ \bibinfo {author}
  {\bibfnamefont {E.~Y.}\ \bibnamefont {Tsymbal}},\ }\href@noop {} {\bibfield
  {journal} {\bibinfo  {journal} {Nat. Commun.}\ }\textbf {\bibinfo {volume}
  {12}},\ \bibinfo {pages} {1} (\bibinfo {year} {2021})}\BibitemShut {NoStop}%
\bibitem [{\citenamefont {\ifmmode~\check{S}\else \v{S}\fi{}mejkal}\ \emph
  {et~al.}(2022)\citenamefont {\ifmmode~\check{S}\else \v{S}\fi{}mejkal},
  \citenamefont {Hellenes}, \citenamefont {Gonz\'alez-Hern\'andez},
  \citenamefont {Sinova},\ and\ \citenamefont
  {Jungwirth}}]{Smejkal2022_PhysRevX_12_011028}%
  \BibitemOpen
  \bibfield  {author} {\bibinfo {author} {\bibfnamefont {L.}~\bibnamefont
  {\ifmmode~\check{S}\else \v{S}\fi{}mejkal}}, \bibinfo {author} {\bibfnamefont
  {A.~B.}\ \bibnamefont {Hellenes}}, \bibinfo {author} {\bibfnamefont
  {R.}~\bibnamefont {Gonz\'alez-Hern\'andez}}, \bibinfo {author} {\bibfnamefont
  {J.}~\bibnamefont {Sinova}}, \ and\ \bibinfo {author} {\bibfnamefont
  {T.}~\bibnamefont {Jungwirth}},\ }\href {\doibase 10.1103/PhysRevX.12.011028}
  {\bibfield  {journal} {\bibinfo  {journal} {Phys. Rev. X}\ }\textbf {\bibinfo
  {volume} {12}},\ \bibinfo {pages} {011028} (\bibinfo {year}
  {2022})}\BibitemShut {NoStop}%
\bibitem [{\citenamefont {Dong}\ \emph {et~al.}(2022)\citenamefont {Dong},
  \citenamefont {Li}, \citenamefont {Gurung}, \citenamefont {Zhu},
  \citenamefont {Zhang}, \citenamefont {Zheng}, \citenamefont {Tsymbal},\ and\
  \citenamefont {Zhang}}]{Dong2022_PhysRevLett_128_197201}%
  \BibitemOpen
  \bibfield  {author} {\bibinfo {author} {\bibfnamefont {J.}~\bibnamefont
  {Dong}}, \bibinfo {author} {\bibfnamefont {X.}~\bibnamefont {Li}}, \bibinfo
  {author} {\bibfnamefont {G.}~\bibnamefont {Gurung}}, \bibinfo {author}
  {\bibfnamefont {M.}~\bibnamefont {Zhu}}, \bibinfo {author} {\bibfnamefont
  {P.}~\bibnamefont {Zhang}}, \bibinfo {author} {\bibfnamefont
  {F.}~\bibnamefont {Zheng}}, \bibinfo {author} {\bibfnamefont {E.~Y.}\
  \bibnamefont {Tsymbal}}, \ and\ \bibinfo {author} {\bibfnamefont
  {J.}~\bibnamefont {Zhang}},\ }\href {\doibase 10.1103/PhysRevLett.128.197201}
  {\bibfield  {journal} {\bibinfo  {journal} {Phys. Rev. Lett.}\ }\textbf
  {\bibinfo {volume} {128}},\ \bibinfo {pages} {197201} (\bibinfo {year}
  {2022})}\BibitemShut {NoStop}%
\bibitem [{\citenamefont {Tsymbal}\ and\ \citenamefont
  {Belashchenko}(2005)}]{Tsymbal2005_JApplPhys_97_10C910}%
  \BibitemOpen
  \bibfield  {author} {\bibinfo {author} {\bibfnamefont {E.~Y.}\ \bibnamefont
  {Tsymbal}}\ and\ \bibinfo {author} {\bibfnamefont {K.~D.}\ \bibnamefont
  {Belashchenko}},\ }\href {\doibase 10.1063/1.1851415} {\bibfield  {journal}
  {\bibinfo  {journal} {J. Appl. Phys.}\ }\textbf {\bibinfo {volume} {97}},\
  \bibinfo {pages} {10C910} (\bibinfo {year} {2005})}\BibitemShut {NoStop}%
\bibitem [{\citenamefont {Tsymbal}\ \emph {et~al.}(2007)\citenamefont
  {Tsymbal}, \citenamefont {Belashchenko}, \citenamefont {Velev}, \citenamefont
  {Jaswal}, \citenamefont {{van Schilfgaarde}}, \citenamefont {Oleynik},\ and\
  \citenamefont {Stewart}}]{Tsymbal2007_ProgMaterSci_52_401}%
  \BibitemOpen
  \bibfield  {author} {\bibinfo {author} {\bibfnamefont {E.~Y.}\ \bibnamefont
  {Tsymbal}}, \bibinfo {author} {\bibfnamefont {K.~D.}\ \bibnamefont
  {Belashchenko}}, \bibinfo {author} {\bibfnamefont {J.~P.}\ \bibnamefont
  {Velev}}, \bibinfo {author} {\bibfnamefont {S.~S.}\ \bibnamefont {Jaswal}},
  \bibinfo {author} {\bibfnamefont {M.}~\bibnamefont {{van Schilfgaarde}}},
  \bibinfo {author} {\bibfnamefont {I.~I.}\ \bibnamefont {Oleynik}}, \ and\
  \bibinfo {author} {\bibfnamefont {D.~A.}\ \bibnamefont {Stewart}},\ }\href
  {\doibase https://doi.org/10.1016/j.pmatsci.2006.10.009} {\bibfield
  {journal} {\bibinfo  {journal} {Prog. Mater. Sci.}\ }\textbf {\bibinfo
  {volume} {52}},\ \bibinfo {pages} {401} (\bibinfo {year} {2007})}\BibitemShut
  {NoStop}%
\bibitem [{\citenamefont {Masuda}\ \emph {et~al.}(2020)\citenamefont {Masuda},
  \citenamefont {Itoh},\ and\ \citenamefont
  {Miura}}]{Masuda2020_PhysRevB_101_144404}%
  \BibitemOpen
  \bibfield  {author} {\bibinfo {author} {\bibfnamefont {K.}~\bibnamefont
  {Masuda}}, \bibinfo {author} {\bibfnamefont {H.}~\bibnamefont {Itoh}}, \ and\
  \bibinfo {author} {\bibfnamefont {Y.}~\bibnamefont {Miura}},\ }\href
  {\doibase 10.1103/PhysRevB.101.144404} {\bibfield  {journal} {\bibinfo
  {journal} {Phys. Rev. B}\ }\textbf {\bibinfo {volume} {101}},\ \bibinfo
  {pages} {144404} (\bibinfo {year} {2020})}\BibitemShut {NoStop}%
\bibitem [{\citenamefont {Masuda}\ \emph
  {et~al.}(2021{\natexlab{a}})\citenamefont {Masuda}, \citenamefont {Itoh},
  \citenamefont {Sonobe}, \citenamefont {Sukegawa}, \citenamefont {Mitani},\
  and\ \citenamefont {Miura}}]{Masuda2021_PhysRevB_103_064427}%
  \BibitemOpen
  \bibfield  {author} {\bibinfo {author} {\bibfnamefont {K.}~\bibnamefont
  {Masuda}}, \bibinfo {author} {\bibfnamefont {H.}~\bibnamefont {Itoh}},
  \bibinfo {author} {\bibfnamefont {Y.}~\bibnamefont {Sonobe}}, \bibinfo
  {author} {\bibfnamefont {H.}~\bibnamefont {Sukegawa}}, \bibinfo {author}
  {\bibfnamefont {S.}~\bibnamefont {Mitani}}, \ and\ \bibinfo {author}
  {\bibfnamefont {Y.}~\bibnamefont {Miura}},\ }\href {\doibase
  10.1103/PhysRevB.103.064427} {\bibfield  {journal} {\bibinfo  {journal}
  {Phys. Rev. B}\ }\textbf {\bibinfo {volume} {103}},\ \bibinfo {pages}
  {064427} (\bibinfo {year} {2021}{\natexlab{a}})}\BibitemShut {NoStop}%
\bibitem [{\citenamefont {Masuda}\ \emph
  {et~al.}(2021{\natexlab{b}})\citenamefont {Masuda}, \citenamefont {Tadano},\
  and\ \citenamefont {Miura}}]{Masuda2021_PhysRevB_104_L180403}%
  \BibitemOpen
  \bibfield  {author} {\bibinfo {author} {\bibfnamefont {K.}~\bibnamefont
  {Masuda}}, \bibinfo {author} {\bibfnamefont {T.}~\bibnamefont {Tadano}}, \
  and\ \bibinfo {author} {\bibfnamefont {Y.}~\bibnamefont {Miura}},\ }\href
  {\doibase 10.1103/PhysRevB.104.L180403} {\bibfield  {journal} {\bibinfo
  {journal} {Phys. Rev. B}\ }\textbf {\bibinfo {volume} {104}},\ \bibinfo
  {pages} {L180403} (\bibinfo {year} {2021}{\natexlab{b}})}\BibitemShut
  {NoStop}%
\bibitem [{\citenamefont {Landauer}(1957)}]{Landauer1957_IBMJResDev_1_3}%
  \BibitemOpen
  \bibfield  {author} {\bibinfo {author} {\bibfnamefont {R.}~\bibnamefont
  {Landauer}},\ }\href {\doibase 10.1147/rd.13.0223} {\bibfield  {journal}
  {\bibinfo  {journal} {IBM J. Res. Dev.}\ }\textbf {\bibinfo {volume} {1}},\
  \bibinfo {pages} {223} (\bibinfo {year} {1957})}\BibitemShut {NoStop}%
\bibitem [{\citenamefont {Landauer}(1970)}]{Landauer1970_PhilMag_21_863}%
  \BibitemOpen
  \bibfield  {author} {\bibinfo {author} {\bibfnamefont {R.}~\bibnamefont
  {Landauer}},\ }\href {\doibase 10.1080/14786437008238472} {\bibfield
  {journal} {\bibinfo  {journal} {Phil Mag.}\ }\textbf {\bibinfo {volume}
  {21}},\ \bibinfo {pages} {863} (\bibinfo {year} {1970})}\BibitemShut
  {NoStop}%
\bibitem [{\citenamefont
  {B\"uttiker}(1986)}]{Buttiker1986_PhysRevLett_57_1761}%
  \BibitemOpen
  \bibfield  {author} {\bibinfo {author} {\bibfnamefont {M.}~\bibnamefont
  {B\"uttiker}},\ }\href {\doibase 10.1103/PhysRevLett.57.1761} {\bibfield
  {journal} {\bibinfo  {journal} {Phys. Rev. Lett.}\ }\textbf {\bibinfo
  {volume} {57}},\ \bibinfo {pages} {1761} (\bibinfo {year}
  {1986})}\BibitemShut {NoStop}%
\bibitem [{\citenamefont
  {B\"uttiker}(1988)}]{Buttiker1988_IBMJResDevelop_32_317}%
  \BibitemOpen
  \bibfield  {author} {\bibinfo {author} {\bibfnamefont {M.}~\bibnamefont
  {B\"uttiker}},\ }\href {\doibase 10.1147/rd.323.0317} {\bibfield  {journal}
  {\bibinfo  {journal} {IBM J. Res. Develop.}\ }\textbf {\bibinfo {volume}
  {32}},\ \bibinfo {pages} {317} (\bibinfo {year} {1988})}\BibitemShut
  {NoStop}%
\bibitem [{\citenamefont {Maekawa}\ and\ \citenamefont
  {Gafvert}(1982)}]{Maekawa1982_IEEETranMagn_18_707}%
  \BibitemOpen
  \bibfield  {author} {\bibinfo {author} {\bibfnamefont {S.}~\bibnamefont
  {Maekawa}}\ and\ \bibinfo {author} {\bibfnamefont {U.}~\bibnamefont
  {Gafvert}},\ }\href {\doibase 10.1109/TMAG.1982.1061834} {\bibfield
  {journal} {\bibinfo  {journal} {IEEE Tran. Magn.}\ }\textbf {\bibinfo
  {volume} {18}},\ \bibinfo {pages} {707} (\bibinfo {year} {1982})}\BibitemShut
  {NoStop}%
\bibitem [{\citenamefont {Groth}\ \emph {et~al.}(2014)\citenamefont {Groth},
  \citenamefont {Wimmer}, \citenamefont {Akhmerov},\ and\ \citenamefont
  {Waintal}}]{Groth2014_NewJPhys_16_063065}%
  \BibitemOpen
  \bibfield  {author} {\bibinfo {author} {\bibfnamefont {C.~W.}\ \bibnamefont
  {Groth}}, \bibinfo {author} {\bibfnamefont {M.}~\bibnamefont {Wimmer}},
  \bibinfo {author} {\bibfnamefont {A.~R.}\ \bibnamefont {Akhmerov}}, \ and\
  \bibinfo {author} {\bibfnamefont {X.}~\bibnamefont {Waintal}},\ }\href
  {\doibase 10.1088/1367-2630/16/6/063065} {\bibfield  {journal} {\bibinfo
  {journal} {New J. Phys.}\ }\textbf {\bibinfo {volume} {16}},\ \bibinfo
  {pages} {063065} (\bibinfo {year} {2014})}\BibitemShut {NoStop}%
\bibitem [{acc()}]{accuracy}%
  \BibitemOpen
  \href@noop {} {}\bibinfo {note} {{ To be precise, the transmissions and the
  LDOS can take so small but finite values, but these nonzero values are due to
  numerics and do not have particular physical meanings, and thus we regard
  these values to be zero throughout this paper.}}\BibitemShut {Stop}%
\bibitem [{\citenamefont {Shim}\ \emph {et~al.}(2008)\citenamefont {Shim},
  \citenamefont {Raman}, \citenamefont {Park}, \citenamefont {Santos},
  \citenamefont {Miao}, \citenamefont {Satpati},\ and\ \citenamefont
  {Moodera}}]{Shim2008_PhysRevLett_100_226603}%
  \BibitemOpen
  \bibfield  {author} {\bibinfo {author} {\bibfnamefont {J.~H.}\ \bibnamefont
  {Shim}}, \bibinfo {author} {\bibfnamefont {K.~V.}\ \bibnamefont {Raman}},
  \bibinfo {author} {\bibfnamefont {Y.~J.}\ \bibnamefont {Park}}, \bibinfo
  {author} {\bibfnamefont {T.~S.}\ \bibnamefont {Santos}}, \bibinfo {author}
  {\bibfnamefont {G.~X.}\ \bibnamefont {Miao}}, \bibinfo {author}
  {\bibfnamefont {B.}~\bibnamefont {Satpati}}, \ and\ \bibinfo {author}
  {\bibfnamefont {J.~S.}\ \bibnamefont {Moodera}},\ }\href {\doibase
  10.1103/PhysRevLett.100.226603} {\bibfield  {journal} {\bibinfo  {journal}
  {Phys. Rev. Lett.}\ }\textbf {\bibinfo {volume} {100}},\ \bibinfo {pages}
  {226603} (\bibinfo {year} {2008})}\BibitemShut {NoStop}%
\bibitem [{eve()}]{evenodd}%
  \BibitemOpen
  \href@noop {} {}\bibinfo {note} {{ Here we write down the equation when $W$
  is even. When $W$ is odd, $W/2$ and $W/2+1$ should be replaced by $(W+1)/2$,
  the center in the $y$-direction. This also applies to
  Eq.~(\ref{eq:ldos_product_center}).}}\BibitemShut {Stop}%
\bibitem [{\citenamefont {Zhang}\ and\ \citenamefont
  {Levy}(1999)}]{Zhang1999_EurPhysJB_10_599}%
  \BibitemOpen
  \bibfield  {author} {\bibinfo {author} {\bibfnamefont {S.}~\bibnamefont
  {Zhang}}\ and\ \bibinfo {author} {\bibfnamefont {P.~M.}\ \bibnamefont
  {Levy}},\ }\href@noop {} {\bibfield  {journal} {\bibinfo  {journal} {Eur.
  Phys. J. B}\ }\textbf {\bibinfo {volume} {10}},\ \bibinfo {pages} {599}
  (\bibinfo {year} {1999})}\BibitemShut {NoStop}%
\bibitem [{\citenamefont {Stanciu}\ \emph {et~al.}(2006)\citenamefont
  {Stanciu}, \citenamefont {Kimel}, \citenamefont {Hansteen}, \citenamefont
  {Tsukamoto}, \citenamefont {Itoh}, \citenamefont {Kirilyuk},\ and\
  \citenamefont {Rasing}}]{Stanciu2006_PhysRevB_73_220402}%
  \BibitemOpen
  \bibfield  {author} {\bibinfo {author} {\bibfnamefont {C.~D.}\ \bibnamefont
  {Stanciu}}, \bibinfo {author} {\bibfnamefont {A.~V.}\ \bibnamefont {Kimel}},
  \bibinfo {author} {\bibfnamefont {F.}~\bibnamefont {Hansteen}}, \bibinfo
  {author} {\bibfnamefont {A.}~\bibnamefont {Tsukamoto}}, \bibinfo {author}
  {\bibfnamefont {A.}~\bibnamefont {Itoh}}, \bibinfo {author} {\bibfnamefont
  {A.}~\bibnamefont {Kirilyuk}}, \ and\ \bibinfo {author} {\bibfnamefont
  {T.}~\bibnamefont {Rasing}},\ }\href {\doibase 10.1103/PhysRevB.73.220402}
  {\bibfield  {journal} {\bibinfo  {journal} {Phys. Rev. B}\ }\textbf {\bibinfo
  {volume} {73}},\ \bibinfo {pages} {220402} (\bibinfo {year}
  {2006})}\BibitemShut {NoStop}%
\bibitem [{\citenamefont {Stanciu}\ \emph {et~al.}(2007)\citenamefont
  {Stanciu}, \citenamefont {Tsukamoto}, \citenamefont {Kimel}, \citenamefont
  {Hansteen}, \citenamefont {Kirilyuk}, \citenamefont {Itoh},\ and\
  \citenamefont {Rasing}}]{Stanciu2007_PhysRevLett_99_217204}%
  \BibitemOpen
  \bibfield  {author} {\bibinfo {author} {\bibfnamefont {C.~D.}\ \bibnamefont
  {Stanciu}}, \bibinfo {author} {\bibfnamefont {A.}~\bibnamefont {Tsukamoto}},
  \bibinfo {author} {\bibfnamefont {A.~V.}\ \bibnamefont {Kimel}}, \bibinfo
  {author} {\bibfnamefont {F.}~\bibnamefont {Hansteen}}, \bibinfo {author}
  {\bibfnamefont {A.}~\bibnamefont {Kirilyuk}}, \bibinfo {author}
  {\bibfnamefont {A.}~\bibnamefont {Itoh}}, \ and\ \bibinfo {author}
  {\bibfnamefont {T.}~\bibnamefont {Rasing}},\ }\href {\doibase
  10.1103/PhysRevLett.99.217204} {\bibfield  {journal} {\bibinfo  {journal}
  {Phys. Rev. Lett.}\ }\textbf {\bibinfo {volume} {99}},\ \bibinfo {pages}
  {217204} (\bibinfo {year} {2007})}\BibitemShut {NoStop}%
\bibitem [{\citenamefont {Radu}\ \emph {et~al.}(2011)\citenamefont {Radu},
  \citenamefont {Vahaplar}, \citenamefont {Stamm}, \citenamefont {Kachel},
  \citenamefont {Pontius}, \citenamefont {D\"{u}rr}, \citenamefont {Ostler},
  \citenamefont {Barker}, \citenamefont {Evans}, \citenamefont {Chantrell},
  \citenamefont {Tsukamoto}, \citenamefont {Itoh}, \citenamefont {Kirilyuk},
  \citenamefont {Rasing},\ and\ \citenamefont
  {Kimel}}]{Radu2011_Nature_472_205}%
  \BibitemOpen
  \bibfield  {author} {\bibinfo {author} {\bibfnamefont {I.}~\bibnamefont
  {Radu}}, \bibinfo {author} {\bibfnamefont {K.}~\bibnamefont {Vahaplar}},
  \bibinfo {author} {\bibfnamefont {C.}~\bibnamefont {Stamm}}, \bibinfo
  {author} {\bibfnamefont {T.}~\bibnamefont {Kachel}}, \bibinfo {author}
  {\bibfnamefont {N.}~\bibnamefont {Pontius}}, \bibinfo {author} {\bibfnamefont
  {H.~A.}\ \bibnamefont {D\"{u}rr}}, \bibinfo {author} {\bibfnamefont {T.~A.}\
  \bibnamefont {Ostler}}, \bibinfo {author} {\bibfnamefont {J.}~\bibnamefont
  {Barker}}, \bibinfo {author} {\bibfnamefont {R.~F.~L.}\ \bibnamefont
  {Evans}}, \bibinfo {author} {\bibfnamefont {R.~W.}\ \bibnamefont
  {Chantrell}}, \bibinfo {author} {\bibfnamefont {A.}~\bibnamefont
  {Tsukamoto}}, \bibinfo {author} {\bibfnamefont {A.}~\bibnamefont {Itoh}},
  \bibinfo {author} {\bibfnamefont {A.}~\bibnamefont {Kirilyuk}}, \bibinfo
  {author} {\bibfnamefont {T.}~\bibnamefont {Rasing}}, \ and\ \bibinfo {author}
  {\bibfnamefont {A.~V.}\ \bibnamefont {Kimel}},\ }\href@noop {} {\bibfield
  {journal} {\bibinfo  {journal} {Nature}\ }\textbf {\bibinfo {volume} {472}},\
  \bibinfo {pages} {205} (\bibinfo {year} {2011})}\BibitemShut {NoStop}%
\bibitem [{\citenamefont {Kurt}\ \emph {et~al.}(2014)\citenamefont {Kurt},
  \citenamefont {Rode}, \citenamefont {Stamenov}, \citenamefont {Venkatesan},
  \citenamefont {Lau}, \citenamefont {Fonda},\ and\ \citenamefont
  {Coey}}]{Kurt2014_PhysRevLett_112_027201}%
  \BibitemOpen
  \bibfield  {author} {\bibinfo {author} {\bibfnamefont {H.}~\bibnamefont
  {Kurt}}, \bibinfo {author} {\bibfnamefont {K.}~\bibnamefont {Rode}}, \bibinfo
  {author} {\bibfnamefont {P.}~\bibnamefont {Stamenov}}, \bibinfo {author}
  {\bibfnamefont {M.}~\bibnamefont {Venkatesan}}, \bibinfo {author}
  {\bibfnamefont {Y.-C.}\ \bibnamefont {Lau}}, \bibinfo {author} {\bibfnamefont
  {E.}~\bibnamefont {Fonda}}, \ and\ \bibinfo {author} {\bibfnamefont
  {J.~M.~D.}\ \bibnamefont {Coey}},\ }\href {\doibase
  10.1103/PhysRevLett.112.027201} {\bibfield  {journal} {\bibinfo  {journal}
  {Phys. Rev. Lett.}\ }\textbf {\bibinfo {volume} {112}},\ \bibinfo {pages}
  {027201} (\bibinfo {year} {2014})}\BibitemShut {NoStop}%
\bibitem [{\citenamefont {Taylor}\ \emph
  {et~al.}(2001{\natexlab{a}})\citenamefont {Taylor}, \citenamefont {Guo},\
  and\ \citenamefont {Wang}}]{Taylor2001_PhysRevB_63_245407}%
  \BibitemOpen
  \bibfield  {author} {\bibinfo {author} {\bibfnamefont {J.}~\bibnamefont
  {Taylor}}, \bibinfo {author} {\bibfnamefont {H.}~\bibnamefont {Guo}}, \ and\
  \bibinfo {author} {\bibfnamefont {J.}~\bibnamefont {Wang}},\ }\href {\doibase
  10.1103/PhysRevB.63.245407} {\bibfield  {journal} {\bibinfo  {journal} {Phys.
  Rev. B}\ }\textbf {\bibinfo {volume} {63}},\ \bibinfo {pages} {245407}
  (\bibinfo {year} {2001}{\natexlab{a}})}\BibitemShut {NoStop}%
\bibitem [{\citenamefont {Taylor}\ \emph
  {et~al.}(2001{\natexlab{b}})\citenamefont {Taylor}, \citenamefont {Guo},\
  and\ \citenamefont {Wang}}]{Taylor2001_PhysRevB_63_121104}%
  \BibitemOpen
  \bibfield  {author} {\bibinfo {author} {\bibfnamefont {J.}~\bibnamefont
  {Taylor}}, \bibinfo {author} {\bibfnamefont {H.}~\bibnamefont {Guo}}, \ and\
  \bibinfo {author} {\bibfnamefont {J.}~\bibnamefont {Wang}},\ }\href {\doibase
  10.1103/PhysRevB.63.121104} {\bibfield  {journal} {\bibinfo  {journal} {Phys.
  Rev. B}\ }\textbf {\bibinfo {volume} {63}},\ \bibinfo {pages} {121104}
  (\bibinfo {year} {2001}{\natexlab{b}})}\BibitemShut {NoStop}%
\bibitem [{\citenamefont {Brandbyge}\ \emph {et~al.}(2002)\citenamefont
  {Brandbyge}, \citenamefont {Mozos}, \citenamefont {Ordej\'on}, \citenamefont
  {Taylor},\ and\ \citenamefont {Stokbro}}]{Brandbyge2002_PhysRevB_65_165401}%
  \BibitemOpen
  \bibfield  {author} {\bibinfo {author} {\bibfnamefont {M.}~\bibnamefont
  {Brandbyge}}, \bibinfo {author} {\bibfnamefont {J.-L.}\ \bibnamefont
  {Mozos}}, \bibinfo {author} {\bibfnamefont {P.}~\bibnamefont {Ordej\'on}},
  \bibinfo {author} {\bibfnamefont {J.}~\bibnamefont {Taylor}}, \ and\ \bibinfo
  {author} {\bibfnamefont {K.}~\bibnamefont {Stokbro}},\ }\href {\doibase
  10.1103/PhysRevB.65.165401} {\bibfield  {journal} {\bibinfo  {journal} {Phys.
  Rev. B}\ }\textbf {\bibinfo {volume} {65}},\ \bibinfo {pages} {165401}
  (\bibinfo {year} {2002})}\BibitemShut {NoStop}%
\bibitem [{\citenamefont {Joon~Choi}\ and\ \citenamefont
  {Ihm}(1999)}]{Choi1999_PhysRevB_59_2267}%
  \BibitemOpen
  \bibfield  {author} {\bibinfo {author} {\bibfnamefont {H.}~\bibnamefont
  {Joon~Choi}}\ and\ \bibinfo {author} {\bibfnamefont {J.}~\bibnamefont
  {Ihm}},\ }\href {\doibase 10.1103/PhysRevB.59.2267} {\bibfield  {journal}
  {\bibinfo  {journal} {Phys. Rev. B}\ }\textbf {\bibinfo {volume} {59}},\
  \bibinfo {pages} {2267} (\bibinfo {year} {1999})}\BibitemShut {NoStop}%
\bibitem [{\citenamefont {Smogunov}\ \emph {et~al.}(2004)\citenamefont
  {Smogunov}, \citenamefont {Dal~Corso},\ and\ \citenamefont
  {Tosatti}}]{Smogunov2004_PhysRevB_70_045417}%
  \BibitemOpen
  \bibfield  {author} {\bibinfo {author} {\bibfnamefont {A.}~\bibnamefont
  {Smogunov}}, \bibinfo {author} {\bibfnamefont {A.}~\bibnamefont {Dal~Corso}},
  \ and\ \bibinfo {author} {\bibfnamefont {E.}~\bibnamefont {Tosatti}},\ }\href
  {\doibase 10.1103/PhysRevB.70.045417} {\bibfield  {journal} {\bibinfo
  {journal} {Phys. Rev. B}\ }\textbf {\bibinfo {volume} {70}},\ \bibinfo
  {pages} {045417} (\bibinfo {year} {2004})}\BibitemShut {NoStop}%
\bibitem [{\citenamefont {Corso}\ and\ \citenamefont
  {Conte}(2005)}]{DalCorso2005_PhysRevB_71_115106}%
  \BibitemOpen
  \bibfield  {author} {\bibinfo {author} {\bibfnamefont {A.~D.}\ \bibnamefont
  {Corso}}\ and\ \bibinfo {author} {\bibfnamefont {A.~M.}\ \bibnamefont
  {Conte}},\ }\href {\doibase 10.1103/PhysRevB.71.115106} {\bibfield  {journal}
  {\bibinfo  {journal} {Phys. Rev. B}\ }\textbf {\bibinfo {volume} {71}},\
  \bibinfo {pages} {115106} (\bibinfo {year} {2005})}\BibitemShut {NoStop}%
\bibitem [{\citenamefont {Dal~Corso}\ \emph {et~al.}(2006)\citenamefont
  {Dal~Corso}, \citenamefont {Smogunov},\ and\ \citenamefont
  {Tosatti}}]{DalCorso2006_PhysRevB_74_045429}%
  \BibitemOpen
  \bibfield  {author} {\bibinfo {author} {\bibfnamefont {A.}~\bibnamefont
  {Dal~Corso}}, \bibinfo {author} {\bibfnamefont {A.}~\bibnamefont {Smogunov}},
  \ and\ \bibinfo {author} {\bibfnamefont {E.}~\bibnamefont {Tosatti}},\ }\href
  {\doibase 10.1103/PhysRevB.74.045429} {\bibfield  {journal} {\bibinfo
  {journal} {Phys. Rev. B}\ }\textbf {\bibinfo {volume} {74}},\ \bibinfo
  {pages} {045429} (\bibinfo {year} {2006})}\BibitemShut {NoStop}%
\end{thebibliography}
\end{document}